\documentclass[10pt,preprint]{aastex}

\newcommand{\teff}{\ifmmode T_{\rm eff} \else $T_{\mathrm{eff}}$\fi} 
\newcommand{\logg}{\ifmmode \log g \else $\log g$\fi} 
\newcommand{\mbol}{\ifmmode M_{\rm bol} \else $M_{\mathrm{bol}}$\fi} 
\newcommand{\lL}{\ifmmode \log \frac{L}{L_{\odot}} \else $\log \frac{L}{L_{\odot}}$\fi}

\newcommand{\msun}{\ifmmode M_{\odot} \else M$_{\odot}$\fi}

\begin{document}

\title{The star formation history of the Milky Way's Nuclear Star Cluster
\footnote{Based on observations collected at the ESO Very Large Telescope (programs 075.B-0547, 076.B-0259, 077.B-0503, 179.B-0261 and 183.B-0100)}} 

\author{
O.~Pfuhl\altaffilmark{a,}\altaffilmark{*},
T.~K.~Fritz\altaffilmark{a},
M.~Zilka\altaffilmark{b},
H.~Maness\altaffilmark{c},
F.~Eisenhauer\altaffilmark{a},
R.~Genzel\altaffilmark{a,c},
S.~Gillessen\altaffilmark{a},  
T.~Ott\altaffilmark{a},  
K.~Dodds-Eden\altaffilmark{a} and
A.~Sternberg\altaffilmark{b}
} 
 
 
 
 \altaffiltext{a} {Max-Planck-Institute for Extraterrestrial Physics, Garching, Germany} 
 \altaffiltext{b} {Sackler School of Physics and Astronomy, Tel Aviv University, Israel} 
 \altaffiltext{c} {Department of Physics, University of California, Berkeley, USA} 
%
%

 \altaffiltext{*} {correspondence: O.~Pfuhl, pfuhl@mpe.mpg.de}

\begin{abstract}
We present spatially resolved imaging and integral field spectroscopy data for 450 cool giant stars within 1\,pc from Sgr\,A*. We use the prominent CO bandheads to derive effective temperatures of individual giants. Additionally we present the deepest spectroscopic observation of the Galactic Center so far, probing the number of B9/A0 main sequence stars ($2.2-2.8\,\msun$) in two deep fields. From spectro-photometry we construct a Hertzsprung-Russell diagram of the red giant population and fit the observed diagram with model populations to derive the star formation history of the nuclear cluster.
 We find that (1) the average nuclear star-formation rate dropped from an initial maximum $\sim10$\,Gyrs ago to a deep minimum 1-2\,Gyrs ago and increased again during the last few hundred Myrs, and (2) that roughly 80\% of the stellar mass formed more than 5\,Gyrs ago;  (3) mass estimates within $\rm R\sim1\,pc$ from Sgr\,A* favor a dominant star formation mode with a 'normal' Chabrier$/$Kroupa initial mass function for the majority of the past star formation in the Galactic Center. The bulk stellar mass seems to have formed under conditions significantly different from the young stellar disks, perhaps because at the time of the formation of the nuclear cluster the massive black hole and its sphere of influence was much smaller than today.

\end{abstract}


\maketitle

\section{Introduction}\label{sec:introduction}
The Milky Way nuclear star cluster (NSC) is of special interest since it is the closest galactic nucleus. It offers the unique possibility to resolve the stellar population and to study the composition and dynamics close to a central black hole at an unrivaled level of detail. 
The ability to resolve individual stars together with continuous monitoring of the innermost stars has proven the existence of a $4.3 \times 10^6 \msun$ supermassive black hole (SMBH) \citep{eisenhauer05,ghez08,gill09} beyond any reasonable doubt. Surveys of galaxies have found scaling relations between the bulge mass and the mass of the central massive object of a galaxy \citep{ferrarese06}. This massive object can be either a NSC or a SMBH, depending on the bulge mass. SMBHs are typically found in massive bulges, while NSCs are common in low mass bulges.  This indicates a mutual evolution of the bulge, the SMBH and the NSC. The Milky Way with its Bulge mass of $10^{10}\,\msun$ falls right in the transition region from galaxies, which are NSC dominated to galaxies, which are SMBH dominated \citep{graham09}. For a critical discussion we refer to \cite{seth08,seth10}. With a radius of 5\,pc and a mass of $2-3\times 10^7 \,\msun$ \citep{launhardt02}, the Milky Way NSC is typical. This lucky coincidence makes the Milky Way's NSC an ideal test case for nuclear co-evolution of galaxies in general. The evolution of a galactic bulge and the central massive object (NSC or SMBH) as well as the physical reason for the scaling relation is however poorly understood. When does a NSC form? Does it evolve simultaneously with the bulge or from a preexisting bulge? How does the SMBH influence the star formation mode? In fact, the age of a NSCs is the largest uncertainty in the determination of its mass \citep{ferrarese06}. Deriving the star-formation history from integrated quantities as obtained from distant galaxies is highly model dependent and is especially uncertain for old ages. These limitations can be overcome in the GC because the stellar populations can be resolved.   \\
Numerous papers have studied the composition of the Milky Way's NSC. These studies have found that the stellar population can be divided into two classes. The cool and evolved giant stars and the hot and young main sequence$/$post-main sequence stars. The existence of massive young stars is evidence for very recent star formation \citep{forr87,allen90}. The most massive stars (WR$/$O stars) reside in a fairly complex structure, which may be described as a combination of a prominent warped disk, a second disk-like structure highly inclined relative to the main disk, and a more isotropic component \citep{paum06,lu09,bartko09} at a projected distance of 0.8-12\arcsec~ from the SMBH Sgr\,A* ($\rm 1\arcsec \equiv 0.04\,pc$, assuming a GC distance of 8.3\,kpc). The GC disk features must have formed in a rapid star burst $\sim 6\rm \,Myrs$ ago \citep{paum06,bartko10}, with a highly unusual initial mass function (IMF) that favored the formation of massive stars ($dN/dm \sim m^\alpha;~ \alpha=-0.45$). This extreme IMF deviates significantly from the standard Chabrier$/$Kroupa IMF with a powerlaw slope of $\alpha=-2.3$ \citep{kroupa01,chabrier03} and seems to exist only in the vicinity of the SMBH. A less massive population of ordinary B-stars can be found in the innermost 1\arcsec, the so called S-stars \citep{eisenhauer05,ghez08,gill09}. The origin of the S-stars is a mystery, because the in-situ formation in the vicinity of the super massive black hole (SMBH) is very improbable. Yet, the K-band luminosity function (KLF) of the S-stars is consistent with a canonical Chabrier$/$Kroupa IMF. The largest population of the resolved stars however, are giants with masses between 1 and 2\,\msun.
Previous kinematic studies of this population have shown that the cluster dynamics are consistent with a relaxed system, slowly rotating in the plane of the galaxy \citep{trippe08,schoedel09}. Detailed abundance determinations of luminous cool giants \citep{carr00,ramirez00,cunha07} found a metallicity distribution peaking at $\rm [Fe/H]=0.14 \pm 0.16$, close to the solar value. \\
The giant population is well suited for observations in the K-band. The giants have typical temperatures between 5100 and 2800\,K. In this temperature range, the CO bandheads at $\rm 2.29 - 2.38\,\mu m$ are the most prominent spectral lines in the K-band. The equivalent widths of these lines correlates with temperature \citep{kle86,wallace96,for00}. This allows the determination of individual temperatures for the red giants (RG). Together with photometric data for the luminosities it is thus possible to construct a Hertzsprung-Russell (H-R) diagram of the GC population and thus to constrain the star formation history in the immediate vicinity of the SMBH.
Detailed studies of the giant population of the Galactic Center have been performed previously by \cite{blum03} and \cite{maness07}. Using a magnitude limited sample of 79 AGB and supergiant stars (50\% complete at $m_{K}<10$) within the central 5\,pc, \cite{blum03} found a variable SFR. They claimed that roughly 75\% of the stars are older than 5\,Gyrs. An intermediate period of low star formation was followed by a recent ($\rm< 100\,Myrs$) period of increased star formation. \cite{maness07} used 329 giant stars ($m_{K}<15.5$), more than 5 magnitudes deeper than \cite{blum03}, albeit at the cost of covering only 12\% of the total area in the central 1\,pc. They found that the giant population is relatively warm, i.e young. Consequently they favored models with either a top-heavy IMF and a constant star formation rate, or with a canonical IMF and an increasing star formation rate during the last few Gyrs. The top-heavy IMF scenario was recently challenged by \cite{loeck10} who argued that this would predict an overabundance of stellar remnants resulting in a mass-to-light ratio larger than observed.\\
The predicted density profile of a relaxed stellar system in the vicinity of a SMBH is a cusp with a radial density profile of $n\sim r^{-\gamma},~\gamma \approx 7/4$ (Bahcall-Wolf cusp). Strictly speaking, this is only true for a single stellar mass cusp. In multiple mass configurations, $\gamma$ can range from $1/2$ to $11/4$. Recent observations have shown that the projected radial distribution of the giant population is actually flat at small radii from Sgr\,A* \citep{do09,buchholz09,bartko10}. There may even be a central hole in the 3D distribution of late type stars \citep{buchholz09,schoedel09}. In either case, the visible distribution seems quite inconsistent with a Bahcall-Wolf cusp. It is possible that a hidden Bahcall-Wolf cusp is present consisting of stellar remnants or unresolved main-sequence stars. However, this would require the stellar luminosity function to change in the few innermost arcseconds. The bright stars need to be removed by mechanisms acting preferentially close to the SMBH. Collisional stripping of stellar envelopes \citep{genzel96,alexander99} is such a mechanism that keeps stars from reaching their peak luminosity. Tidal stripping close to the SMBH can deplete the giants in a similar way \citep{davies06}. Another explanation might be a top-heavy or truncated IMF \citep{nayakshin05} that produces preferentially massive stars as observed in the young disks. Massive stars quickly become stellar remnants compared to evolved low mass stars that dominate the KLF. \cite{merritt09} argues that the absence of a Bahcall-Wolf cusp can be naturally explained if the stellar population is not relaxed. They find that the relaxation time scale in the innermost parsec is greater than $5-10\rm\,Gyrs$. The most important question to discriminate between depletion processes and a relaxation time effect is therefore, whether the system is old enough to be relaxed. \\

The paper is organized as follows; In section \ref{sec:obs_and_proc} we present the data and in section \ref{sec:spec_class} the spectral classification and calibration of spectral indices. In section \ref{sec:const_HRdiag} we construct the H-R diagram that is then fit with model populations in section \ref{sec:SFR_fit}. The star formation history, IMF and mass composition of the nuclear cluster is presented in section \ref{sec:results}. The results are discussed and compared with other works in section \ref{sec:discussion}. We conclude in section \ref{sec:conclusion}.


\section{Observations and data processing}\label{sec:obs_and_proc}
This work relies on spectroscopic and imaging data obtained at the VLT in Cerro Paranal Chile between 2003 and 2010. The observations were carried out under the program-ids 075.B-0547, 076.B-0259, 077.B-0503, 179.B-0261 and 183.B-0100.
\subsection{Imaging and photometry}
The photometric data were obtained with the adaptive optics camera NACO \citep{rousset03, hartung03}. The photometric reference images were taken on the 29th of April 2006 and on the 31st of March 2010. We used the H- and K'-band filter together with a pixel scale of 27\,$\rm{mas/pixel}$. To each image we applied sky-subtraction, bad-pixel and flat-field correction \citep{trippe08}. All images of good quality obtained during the same night were then combined to a mosaic with a FoV of $40\times40\arcsec$. 
\subsubsection{Diffuse background}\label{sec:diffuse_background}
In terms of mass, the resolved stellar population represents only the tip of the iceberg. The bulk of the stellar mass is unresolved. Therefore the diffuse background emission of the GC contains valuable information on the cluster composition and its formation.  Naturally it is very challenging to estimate the unresolved background in a crowded stellar field. A significant fraction of the background light originates from the uncorrected seeing halos of bright stars. Furthermore, anisoplanatism complicates precise photometry across the Field of View (FoV) when using AO. By using multiple point spread function (PSF) photometry on NACO images \cite{schoedel10b} tried to overcome these limitations. The resulting photometry for resolved stars was published in \cite{schoedel10a} and a map containing the unresolved background flux was published in \cite{schoedel10b}. We used H-band data from both publications to derive the fraction of light contained in the diffuse background. The H-band has the advantage that it is least affected by the surrounding nebular emission. \\
Since no completeness map was published by  \cite{schoedel10b}, we used a superb H-band mosaic with Strehl $>20\%$ from the 31st of March 2010. We applied StarFinder \citep{diolaiti00} and converted our detections in magnitudes by referencing them to Table\,A2 of \cite{schoedel10a}. To derive the completeness, we applied a common technique of creating and re-detecting artificial stars of various brightness. As a limiting magnitude, separating diffuse background from resolved sources, we used $m_{H,\rm cut}=19.45 \pm 0.12$. The magnitude was chosen such that the total uncorrected flux of all detected stars is equal to the completeness corrected flux of stars brighter than the magnitude cut. Using an extinction of $A_H=4.65 \pm0.12$ \citep{fritz11} and a GC distance of $R_0=8.3\pm0.35$\,kpc \citep{ghez08,gill09} we obtained an absolute separating magnitude of $M_{H,{\rm cut}}=0.27 \pm0.34$. The error estimate accounts for the absolute photometric uncertainty, the cut-off, extinction and distance uncertainties. We then subtracted the difference between the stellar flux (up to the limiting magnitude) in our H-band image and the stellar flux in the \cite{schoedel10a} list from the diffuse background flux. Furthermore, we subtracted the known early-type stars since they dominate the light in the inner 10\arcsec. \\ 
We find that the diffuse background ($m_{H}>19.45 \pm 0.12$) contains $H_{\rm diff}/H=27\pm 9\%$ of the total (diffuse + resolved) H-band flux. The $1\,\sigma$ error contains the variation across the field (6\%) and the uncertainty due to the cutoff between diffuse and resolved light $M_{ H,\rm cut}$ (2\%). Taking into account the completeness, we added another 3\% systematic error due to undetected early-type stars.
\subsubsection{Mass-to-light ratio}
To derive the total H-band luminosity of the inner 1.2\,pc, we used the resolved and unresolved stellar H-band flux stated in \cite{schoedel10a,schoedel10b}. To account for the spatially varying extinction of the GC, we used their published extinction map. Unfortunately the extinction map is only available for K-band. Therefore we scaled the map to an average H-band extinction of $A_H=4.65$ \citep{fritz11}. Correcting for extinction and taking the distance into account, we obtained a luminosity of $H=4.51\pm0.54\times10^6 \,L_{H,\odot}$ enclosed within a projected radius of $\rm r_{2D}<1.2\,pc$. From this we subtracted the contribution of known early-type stars ($0.64\times10^6 \,L_{H,\odot}$). While the young stars are known to be close in 3D to the center, only $50\pm5\%$ of the old stars at a projected distance of $\rm r_{2D}<1.2\,pc$ are also contained within a 3D distance of 1.2\,pc (assuming the radial density profile of Sch\"odel et al.2007). Applying both corrections we find a total luminosity of $H(R<1.2{\rm \,pc})=1.94\pm0.33\times10^6 \,L_{\odot,H} $. Of this value $0.52 \pm0.2 \times10^6 \,L_{\odot,H}$ can be attributed to the diffuse background. The dynamical mass enclosed (excluding the SMBH) is $M(R<1.2{\rm \,pc})=1.4\pm0.7\times10^6 \,\msun$ \citep{genzel10}. Thus we derive a total mass-to-light ratio of $M/H=0.7\pm0.4 \,\msun/L_{\odot,H}$ and a diffuse mass-to-light ratio of $M/H_{\rm diff}=2.6\pm1.7\, \msun/L_{\odot,H}$. \\
By assuming an average intrinsic colour of $m_{H-K}=0.2$ we can convert the total H-band luminosity into a K-band luminosity of $K(R<1.2{\rm \,pc})=3.0\pm0.44\times10^6 \,L_{\odot,K}$. The corresponding mass-to-light ratio is $M/K=0.5\pm0.3 \,\msun/L_{\odot,K}$. The mass-to-diffuse-light ratio is $M/K_{\rm diff}=1.9\pm1.2 \,\msun/L_{\odot,K}$. The bolometric mass-to-light ratio is $M/L_{\rm bol}=0.7\pm0.4 \,\msun/L_{\odot,\rm bol}$.
\subsection{Spectroscopy}
Our spectroscopic data were obtained with the adaptive optics assisted integral field spectrograph SINFONI \citep{eis03, bon04}. We used all high quality SINFONI data sets available to us, and taken between 2003 and 2010. In total we used 32 fields with pixel scales between 25 and 100\,mas. The data output of SINFONI consists of cubes with two spatial axes and one spectral axis. Depending on the plate scale, an individual cube covers $0.8\arcsec \times0.8\arcsec$ or $3.2\arcsec \times3.2\arcsec$; the spectral resolution varies between 2000 and 4000 depending on the chosen bandpass and the field of view. Roughly 70\% of the stars used in this work were observed at a resolution of $R\sim2000$. We used the data reduction SPRED \citep{schreiber04,abu06}, including bad-pixel correction, flat-fielding and sky subtraction. The wavelength scale was calibrated with emission line gas lamps and fine-tuned on the atmospheric OH lines. Finally we removed the atmospheric absorption features by dividing the spectra through a telluric spectrum obtained in the respective night.
\subsubsection{Deep spectroscopy of the GC}
\begin{figure}
\begin{center}
\includegraphics[width=\columnwidth]{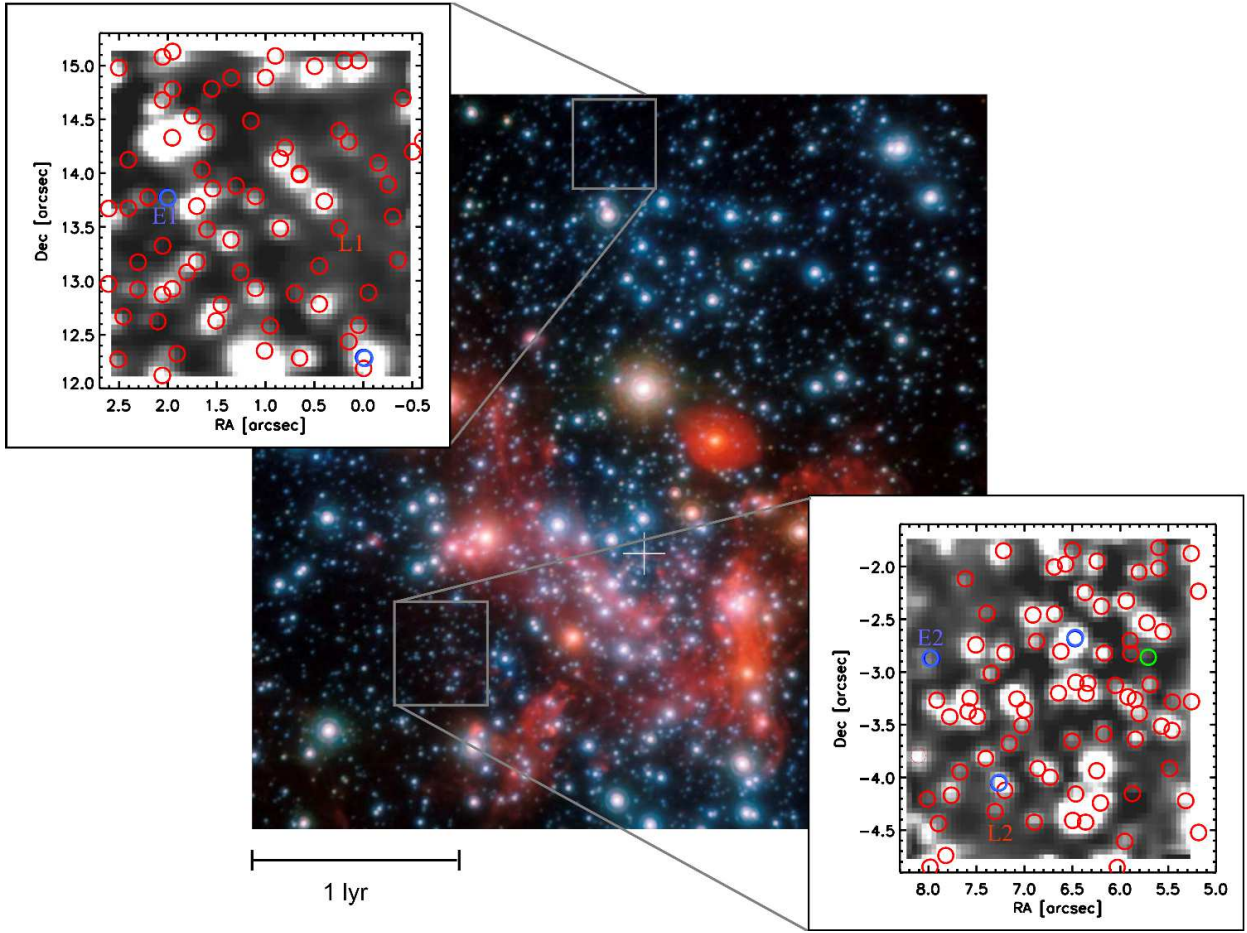}
\caption{Three-color image of the innermost 1\,pc of the Milky Way (K-band: blue; L-band: red). The white cross indicates the position of Sgr\,A*. The U-shaped nebular emission visible in the L-band, the so called Mini-Spiral, is powered by the ionizing radiation of the young O$/$WR stars in the vicinity. The position of the two deep spectroscopic fields is indicated. All photometrically detected stars not confused with neighboring stars are indicated by open circles. The color coding resembles the stellar type. Cold giants with confirmed CO bandheads are indicated in red. Hot main sequence stars identified by $\rm{Br}\gamma$ absorption are indicated in blue. Green stars are not identified.}
\label{fig:overview}
\end{center}
\end{figure}

Very good seeing conditions and long integration times on two SINFONI fields allowed us to perform the deepest census of the GC population so far. The two fields cover roughly 18 square arcseconds and are $\sim50\%$ complete down to $m_K < 18$.
The location of the fields can be seen in Fig.\,\ref{fig:overview}.  Both fields were placed in regions with low extinction and without bright stars in the vicinity. One of the fields (East), at a distance of $7.4\arcsec$, probes the disk region. The second field (North) is located at the outer rim of the disks at $13.5\arcsec$. The richness of faint stars can be seen in Fig.\,\ref{fig:overview}. The identification of a $m_K < 17$ star as late-type requires only two hours of integration on source. The prominent CO bandheads are easy to identify and the multiplicity of the lines helps to rule out misidentifications in noisy spectra. However, the identification of an early-type star is much more challenging. The early-type stars are identified by the hydrogen $\rm Br\gamma$ line. Unfortunately the line is less prominent than the CO bandheads. The identification is further hindered by the surrounding nebular $\rm Br\gamma$ emission. This emission is patchy and varies significantly in intensity and velocity on scales of 1\arcsec. Depending on the location of the star, this can lead to a wrong background subtraction during the data analysis mocking a stellar line. Because the stellar absorption line in A-stars is a broad Lorentzian, it is clearly different from the narrow Gaussian emission line of the gaseous background. Thus in case of sufficient SNR an identification is possible.  To achieve a sufficient signal-to-noise ratio (SNR) an integration time of about five hours on source is required to identify an early-type star as such with $m_K < 18$.
 
\subsection{Source selection}
The selection of stars used in the H-R diagram fitting is based on a master list of $\sim6000$ stars at projected radii between $0.1\arcsec <r< 25\arcsec$ (or $4\times 10^{-3}$ to 1\,pc), for which \cite{trippe08} derived proper motions as well as K-band photometry. The list consists of well isolated stars identified on several images without overlapping neighbors ($\rm{separation > 130\,mas}$). For our final analysis we retained only those entries in the master list that also had an unambiguous identification in a corresponding SINFONI cube. In total we collected $\rm{\sim1300}$ spectra. After removing duplicates (due to overlapping fields), roughly 1000 spectra remained. Out of this sample we used only stars with a $\rm{SNR >10}$ and a CO index $\rm EW(CO)>3.5$ (i.e. no early-type stars). The SNR was measured in the continuum bands stated in Table\,\ref{tab:CO}. This left us with about 800 giants for which we were able to determine individual temperatures and velocities. In order to have a homogeneous and deep sample, we used in the H-R diagram fit only stars contained in fields with a completeness $>50\%$ at ${m_K=15}$ (see next section). The remaining sample consisted of 450 stars with magnitudes between ${0.5 < M_{K} < -8}$. The contribution of fore- and background stars is negligible in the GC. The contamination due to stars which are not members of the cluster is of the order 1\% \citep{buchholz09}. 

\begin{deluxetable}{lc}
\tablecaption{Equivalent width measurement intervals\label{tab:CO}}
\tablewidth{0pt}
\tablehead{
\colhead{Feature} &
\colhead{Wavelengths ($\rm \mu m$)}
}
\startdata
$\rm ^{12}CO(2,0)$ band & 2.2910-2.3020 \\ 
$\rm ^{12}CO(2,0)$ continuum I & 2.2300-2.2370 \\
$\rm ^{12}CO(2,0)$ continuum II & 2.2420-2.2580 \\
$\rm ^{12}CO(2,0)$ continuum III & 2.2680-2.2790 \\
$\rm ^{12}CO(2,0)$ continuum IV & 2.2840-2.2910 \\
\enddata
\end{deluxetable}

\smallskip

\subsection{Detection probability}
\begin{figure}
\begin{center}
\includegraphics[width=0.8\columnwidth]{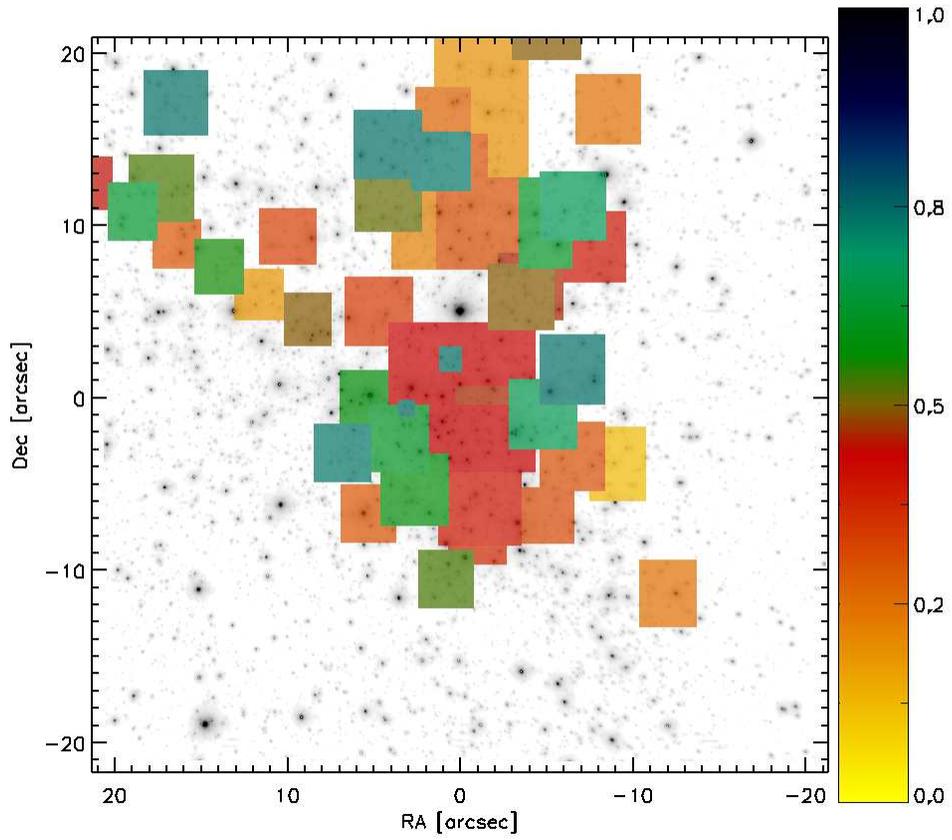}

\caption{The AO assisted SINFONI fields overlaid on a GC image. The color of the fields indicates the combined spectroscopic and photometric completeness for a $m_K=15$ star. The completeness is mainly driven by the exposure time of the individual fields and the observing conditions of the respective night.
}
\label{fig:SINFONI_compl}
\end{center}
\end{figure}

The spectroscopic fields used in this work differ in spatial resolution, covered area and integration time. The main limitation for spectroscopy in the GC is stellar crowding. This is why we probed most of the inner 1\,pc with the small scale (100, 25\,mas) AO-assisted modes of SINFONI. For these observations, we used typically one hour of integration time on source. Two additional cubes (100\,mas) are exceptionally deep with integration times of $4.5$ and $7.8$ hours on source. The spectroscopic completeness of these cubes is $>50\%$ at $m_K<18$. The results of the deep observations are discussed in Sec.\,\ref{sec:ultradeep}.  The spectroscopic completeness was determined by comparing the total number of stars contained in the master list, with the number of stars for which we could extract spectra. The photometric completeness of the images was determined as described in Sec.\,\ref{sec:diffuse_background}. All observed fields and the combined completeness (for $m_{K}=15$) are shown in Fig.\,\ref{fig:SINFONI_compl}. Fields with a high completeness of up to 80\% are found at several arcsecond distance from Sgr\,A*. Further in, the completeness degrades due to stellar crowding. Only the inner one arcsecond is sampled with the smallest pixel scale (25\,mas) and reaches a completeness of about 50\%.

\section{Spectral classification}\label{sec:spec_class}
The K-band provides two prominent spectral features allowing a spectral classification. Early-type stars ($\rm T>5000\,K$) can be identified by the presence of a $\rm Br\gamma$ absorption line. Late-type stars ($\rm T<5000\,K$) on the other hand show weak or no $\rm Br\gamma$ absorption but strong absorption line blends known as CO bandheads. The bandheads appear at the spectral type $\rm \sim G4$ and increase in strength up to M7. The strength also increases with luminosity class from dwarfs to giants and to supergiants. Yet, for this work only giants are of interest. Main sequence stars with spectral type G4 and cooler are fainter ($m_{K} > 20$) than the detection limit. Supergiants of the same spectral type on the other hand are extremely bright ($m_{K} < 9$) and easy to identify \citep{blum03}. 
\subsection{Deep census of the GC population}\label{sec:ultradeep}
The two deep SINFONI fields allowed a deep census of the GC population with 50\% completeness down to $m_K<18$.
\subsubsection{A-star detection}\label{sec:Astar_detection}
\begin{figure}
\begin{center}
\includegraphics[width=0.8\columnwidth]{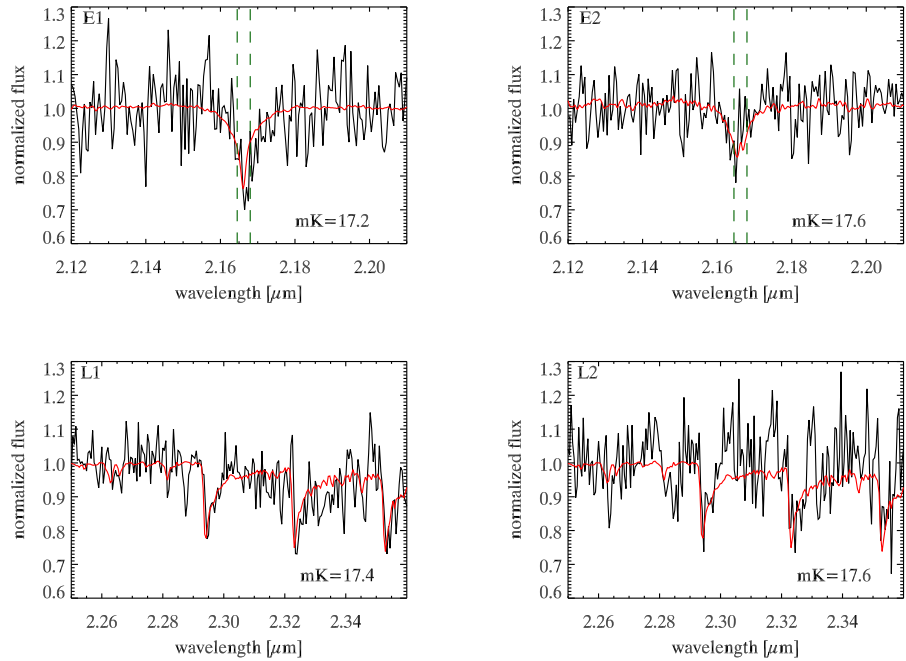}
\caption{Spectra of the faint $\rm A0/B9V$ main-sequence stars in the Galactic Center (upper panels). The red line shows $\rm B9/A0\,V$ template spectra of the solar neighborhood \citep{wallace97}. The dashed bars indicate the width of the nebular emission in the vicinity of the stars. The name in the upper left corner relates the spectrum to the star in Fig.\,\ref{fig:overview}. For comparison, two late-type spectra of similar brightness ares shown in the lower panels. The corresponding late-type templates are indicated in red.}
\label{fig:faint_spectra}
\end{center}
\end{figure}

Fig.\,\ref{fig:overview} shows the deep fields. Red and blue circles mark stars identified as late-type or early-type. One star was not confirmed as early-type (green circle), but showed no CO bandheads and is therefore deemed to be an early-type candidate.
 Within the 18 square arcseconds of the two fields, we thus detected five early-type stars and one candidate. Their spectral type was determined by comparison with template spectra of \cite{wallace97} and the absolute K-band magnitude $M_K$ taken from \cite{cox00}. Three of them were brighter than $m_K < 17$ and were identified as dwarfs with spectral types between B2 and B8. The two faintest early type stars identified were B9$/$A0 dwarfs with $m_K = 17.2$ and $m_K = 17.6$. These stars are the faintest main-sequence stars reported so far in the Galactic Center. Stars of that spectral type have main-sequence lifetimes between 360-730\,Myrs and masses between $2.2-2.8\,\msun$ \citep{cox00}. The spectra of the two A-stars are shown in Fig.\,\ref{fig:faint_spectra}. For comparison, two spectra of late-type giants of equal K-band luminosity are shown. The candidate is of similar luminosity falls in the same main-sequence category.

\subsubsection{K-band luminosity function}
\begin{figure}
\begin{center}
\includegraphics[width=0.6\columnwidth]{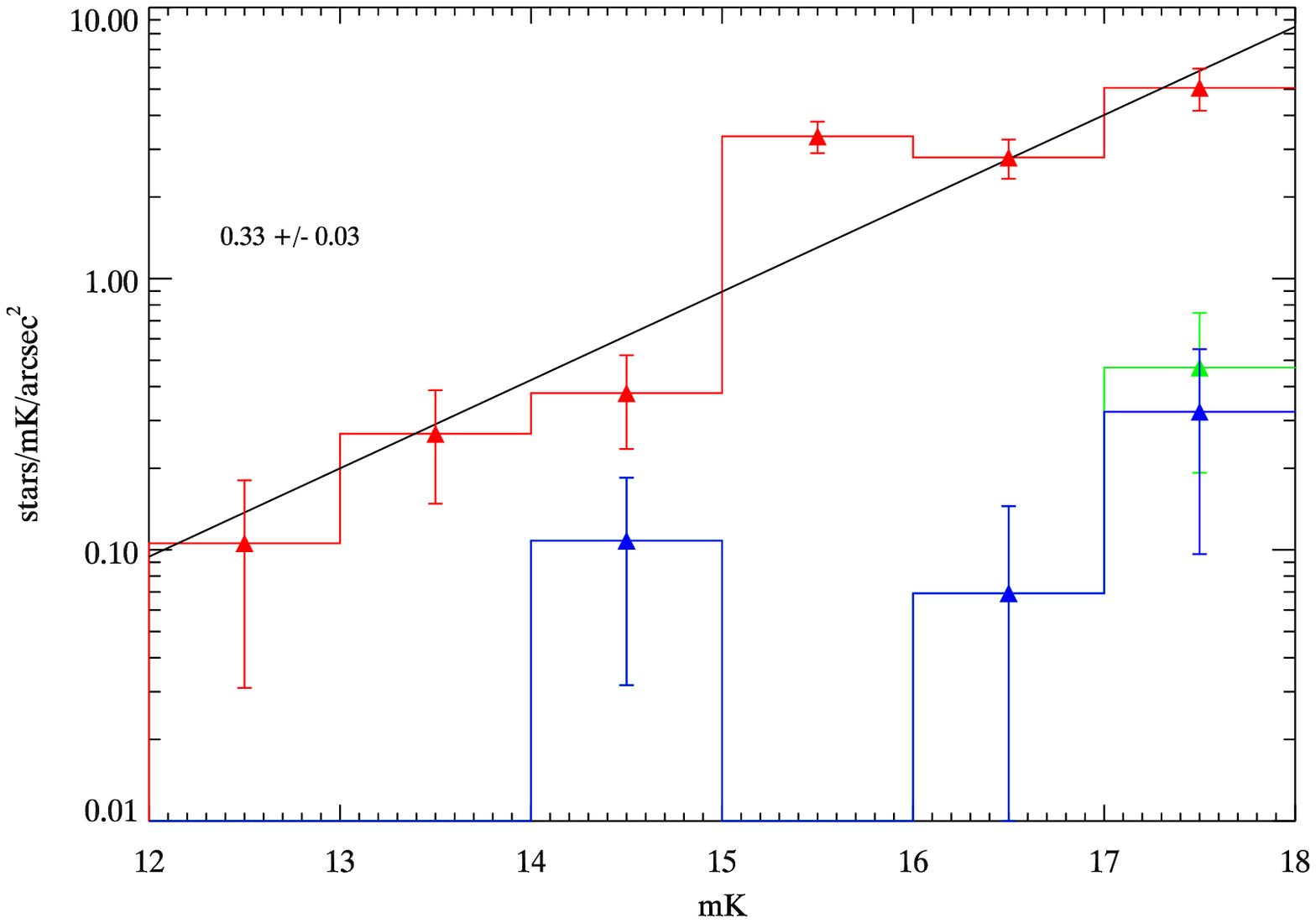}
\caption{Completeness corrected KLF of the confirmed early- (blue) and late-type (red) stars found within the two deep spectroscopic fields. The green KLF shows the early type stars including the unidentified star. The solid line indicates the best-fit powerlaw to the late-type KLF.
}
\label{fig:KLF}
\end{center}
\end{figure}

The deep census of the GC allowed us to construct a KLF with $12<m_K<18$ (Fig.\,\ref{fig:KLF}). The best fitting slope of the luminosity distribution ${\rm d\, log}N/{\rm d}m_K=\beta$ is found to be $\beta = 0.33 \pm 0.03$. This is consistent with the results of \cite{buchholz09}, who determined the slope of stars with a limiting magnitude of $m_K < 15.5$. As already discussed by \cite{alexander99b,genzel03,buchholz09}, the slope is typical for an old, bulge-like population. This is in good agreement with our H-R diagram fitting (Sec.\,\ref{sec:fit}) showing that most of the star formation occurred more than 5\,Gyrs ago. However, the KLF alone is not sufficient to constrain the star formation history for an old population. The KLF alone cannot constrain the actual formation history and the IMF since the KLF slope is very insensitive to both parameters \citep{loeck10}. For example, the admixture of young and bright giants is hardly detectable in the KLF due to their low number. Yet, they can be easily identified in the H-R diagram. 

\subsubsection{Origin: Disk members?}
One of the observed fields is centered on the disk region, while the second field probes the outer rim of the disks. Thus the early-type stars in the two observed fields are potential disk members. To test for membership, we used the same method as described in \cite{bartko10}. The 3D velocity and the positions of all five stars are given in Table\,\ref{tab:early}. The last column indicates a likely disk membership. We find that the two A-stars are clearly not disk members. Among the brighter B-stars, only one is consistent with the counter-clockwise disk (CCWS). This supports the results of \cite{bartko10} who find a disk IMF skewed towards massive stars. Their disk KLF predicts roughly the same number of stars with $m_K=14$ as with $m_K=17$. A Chabrier$/$Kroupa IMF on the other hand predicts five times more $m_K=17$ stars. Although the statistical significance of only five early-type stars is limited, it is intriguing that we find one $m_K=14$ disk member (consistent with the average B-star density) but no fainter disk members. The faint B$/$A-stars exhibit an isotropic orientation \citep{bartko10} and are most likely the remains of older star bursts.

\begin{table}[t!] 
\begin{center}
 \caption{Early type stars contained in the deep spectroscopic fields. \label{tab:early}} 
\begin{tabular}{cccccccc} 
\hline\hline 
Field & mK & RA\tablenotemark{a} [\arcsec] & Dec\tablenotemark{a} [\arcsec] & $\rm v_x ~[mas/yr]$ & $\rm v_y ~[mas/yr]$ &  $\rm v_z ~[km/s]$ & Disk member\tablenotemark{b} \\ 
\hline    
North & 14.1 & 0.0 & 12.3 & $2.44\pm 0.10$ & $0.55 \pm 0.24$ & $-100 \pm 40$ & CCWS   \\
North & 17.2 &2.0 & 13.8 & $0.62\pm 0.27$ & $0.96 \pm 0.4$ & $116 \pm 55$  & no \\
East & 14.2 & 6.5 & -2.7 & $-2.73\pm 0.14$ & $-3.73\pm 0.26$ & $-60 \pm 7$  & no \\
East & 16.4 & 7.3 & -4.1 & $2.89\pm 0.23$ & $ -1.25 \pm 0.16$ & $-163 \pm 44$ & no  \\
East & 17.7 & 8.0 & -2.7 & $2.19\pm 0.39$ & $-0.12\pm 0.24$ & $-41 \pm 52$  & no \\
\hline 
\end{tabular} 
\tablenotetext{a}{Distance in arcseconds relative to Sgr\,A*.}
\tablenotetext{b}{Potential members of the clockwise disk (CWS) or the counter-clockwise disk (CCWS) as defined by \cite{bartko09}.}
\end{center}
\end{table}

\subsection{Supergiant IRS7}
Two red supergiants are known to reside within 1\,pc of Sgr\,A* \citep{blum03, paum06}. Those supergiants are significantly younger than the rest of the old stellar population. The supergiant IRS\,7 is the youngest of this particular luminosity class. \cite{carr00} find an initial mass of 20\,\msun~ and stellar age of $\rm\sim 7\,Myrs$ and attribute it to the recent starburst. This is supported by its kinematics (Bartko; priv. comm.) placing IRS\,7 on the clockwise disk system. We simulated a 6\,Myr star burst as found by \cite{paum06} and \cite{bartko10} with the IAC-STAR code (see following sections). The predicted ratio of red supergiants ($\teff< 5000\,{\rm K};~ m_K < 9$) to blue supergiants ($\teff > 6000\,{\rm K};~ 9<m_K<14$) is between 0.4-0.5\% (independent of the assumed IMF). This is in good agreement with the observed ratio of 118 blue supergiants with $ m_K<14$ \citep{bartko10} to one supergiant, i.e. a ratio of 0.9\%. The second supergiant has an age of a few ten Myrs \citep{blum03} and is thus older than the disks.
\subsection{CO index definition}\label{sec:CO_index}
\begin{figure}
\begin{center}
\includegraphics[width=0.45\columnwidth]{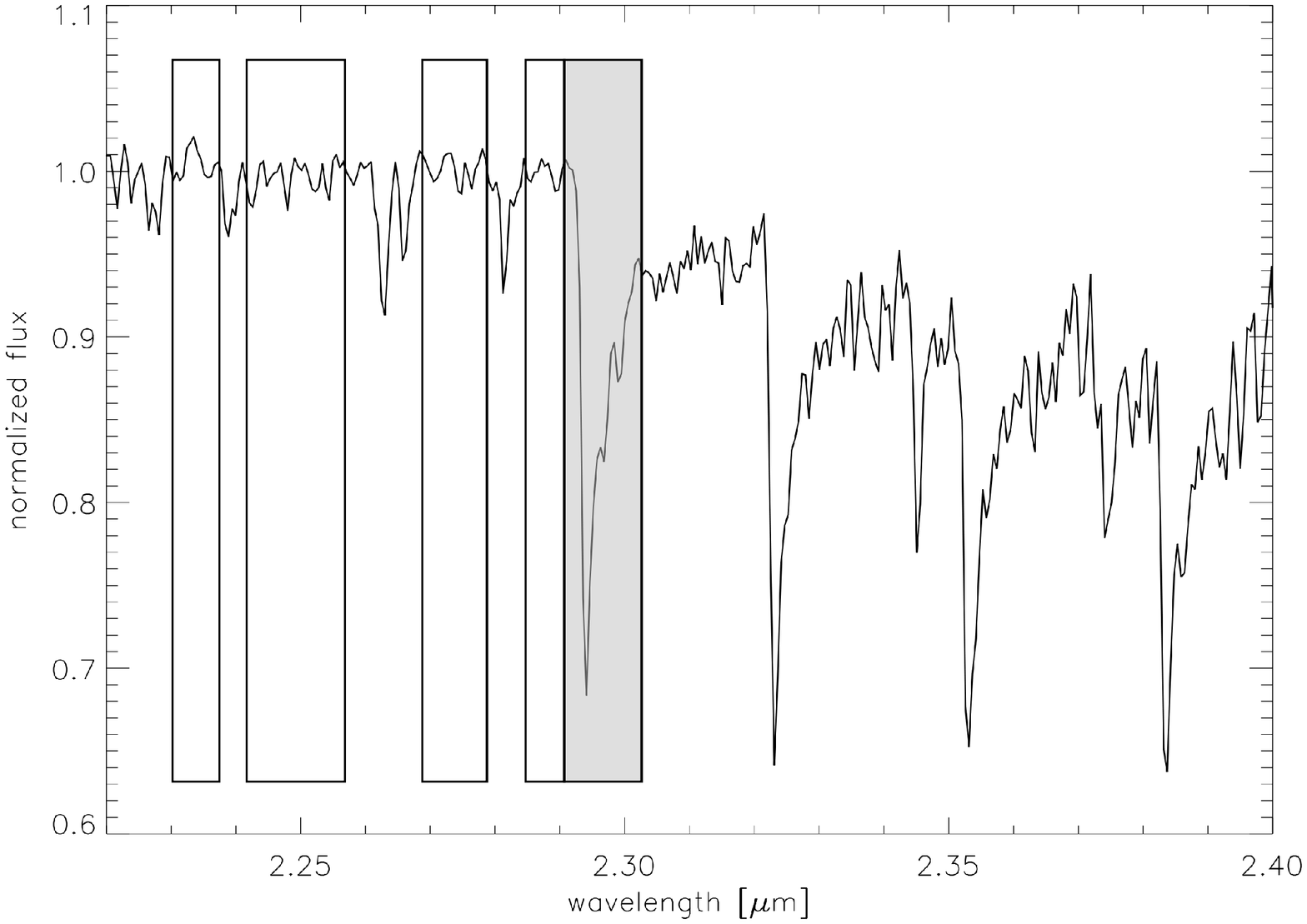}
\includegraphics[width=0.45\columnwidth]{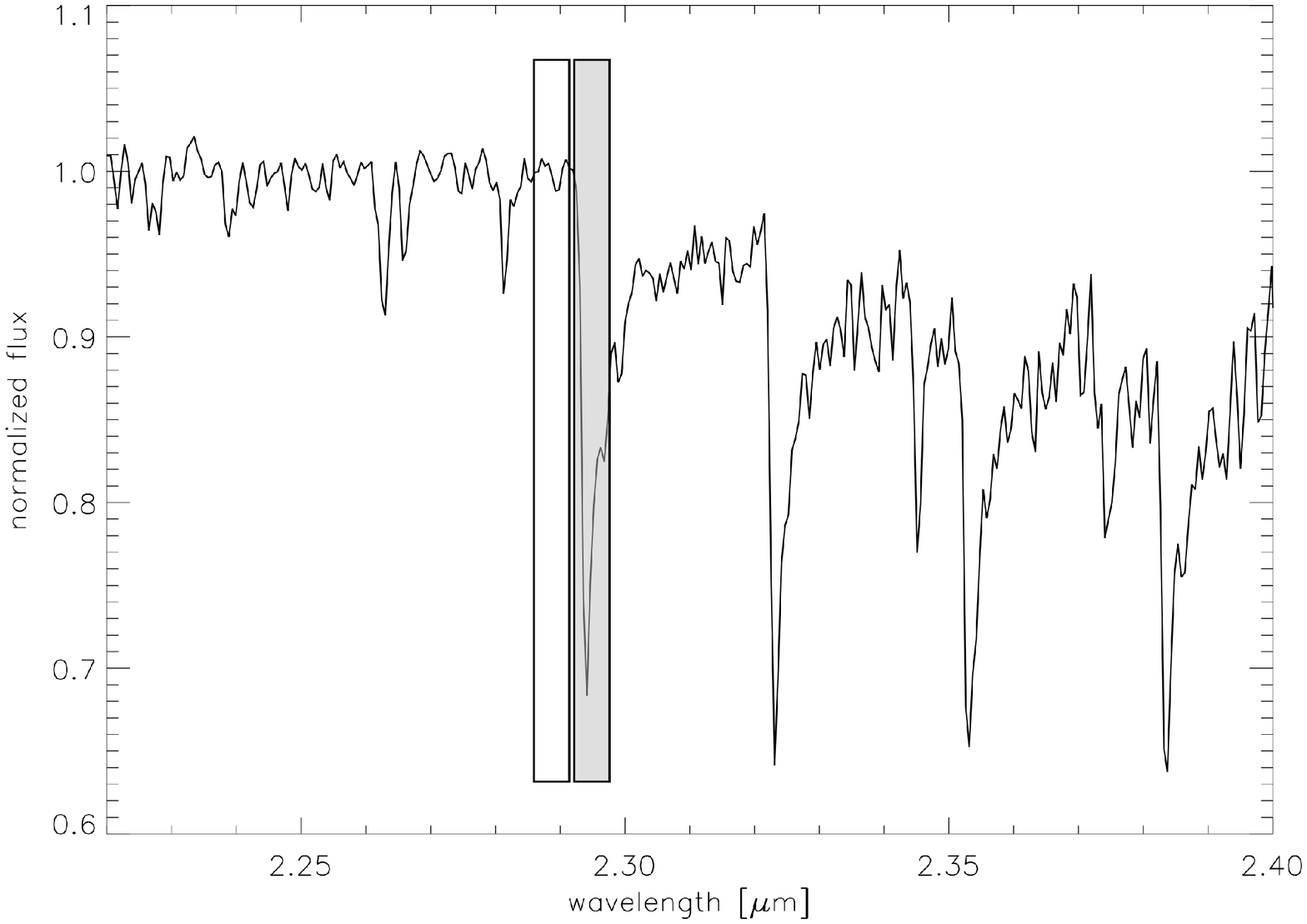}
\caption{Left: Index definition of \cite{frogel01}. The gray area indicates the line region and the white area indicates the continuum regions used for the fit. Right: Index definition of \cite{kle86}. An adaptation of that index was used in \cite{blum03} and \cite{maness07}.
}
\label{fig:index_def}
\end{center}
\end{figure}

The $\rm ^{12}CO(2,0)$ bandhead has been widely used as a temperature indicator. This has brought up numerous definitions of CO indices. A detailed comparison of regularly used index definitions was recently performed by \cite{marmol08}. The analysis showed that some index definitions are systematically more affected by spectral resolution, velocity error, curvature of the spectrum and SNR than other definitions. In particular, the index definition of Kleinman\,\&\,Hall 1986 proved to be very sensitive to those effects. The index measures the continuum and the line flux in two narrow bands in close proximity.A similar index was used in previous studies of the Galactic Center population by \cite{maness07} and \cite{blum03}. However the authors used a wider bandpass than \cite{kle86} ($0.015\,\mu m$ instead of $0.0052\mu m$). In the following the index is referred to as (BL03). Systematic errors in the measurement of the CO strength lead to systematic errors in the temperature estimation. Since the temperature is a tracer of the stellar age, this might lead to a bias in the determination of the star formation history. Indeed, the paucity of cool stars in the GC is the chief constraint which led \cite{maness07} to conclude a top-heavy IMF may have persisted throughout the star formation history of the Galactic Center. To estimate the impact of systematics in light of this recent work, we tested the index BL03 against various effects and compared its performance with the alternate index proposed by Frogel et al.\,(2001, FR01). It uses several narrow bandpasses to estimate the CO continuum with a linear fit (see Tab.\,\ref{tab:CO}). The line and continuum regions for both index definitions can be seen in Fig.\,\ref{fig:index_def}.

\subsubsection{Index computation}
Before computing the indices, each stellar spectrum was shifted to rest-wavelength. We then divided the spectrum by a second-order polynomial fit to remove the curvature of the spectrum. For the continuum fit, we excluded regions with significant absorption lines; the polynomial fit is necessary to account for the the large ($A_K\sim3$) and spatially variable extinction in the GC. We then computed the BL03 index according to the recipe of \cite{blum03}. The FR01 index was computed in a similar way, with the bandpasses as described in \cite{frogel01}. The continuum level $w_{\rm C}(\lambda)$ is estimated with a linear fit to the intervals stated in Table\,\ref{tab:CO}. The equivalent width is measured according to: 
\begin{equation}
{\rm EW(CO)}=\int_{\rm{band}} (\frac{w_{\rm C} - w_{\rm{line}}}{w_{\rm C}})\rm d\lambda.
\end{equation}
We tested both indices (BL03, FR01) with template spectra of different resolution. We also reddened the template spectra artificially to determine the impact of extinction. 
\subsubsection{Systematic error sources}
Reducing the resolution from $\rm R\sim3000$ to $\rm R\sim2000$ decreases the BL03 CO index by a factor of $\approx 0.92$. The artificial reddening and the corresponding change of curvature of the spectrum, causes a reduction of $\approx 0.94$. Both effects cause the BL03 CO index to be underestimated by a factor between 0.90-0.85. The systematic underestimation of the BL03 CO strength leads to an overestimation of the stellar temperatures by $\rm \sim 200\,K$. The FR01 index decreased by less than a factor $0.98$ due to extinction. Degrading the resolution from $\rm R\sim3000$ to $\rm R\sim2000$ shows no measurable impact. We estimate the combined systematic effect to be less than $0.97~ (50\,K)$. The BL03 index suffers also from contamination of the Mg\,I line contained in the continuum bandpass of the index. The average Mg\,I line strength of the GC giants is similar to the calibration giants. Thus, the continuum estimation is not biased. Yet, the line shows some intrinsic scatter introducing a statistical error in the CO index computation. To minimize the statistical and systematic error sources, we therefore adopted the index definition of FR01.
We have to note that the results of \cite{blum03} did not suffer from the resolution dependence of their CO index 
because the bulk of their GC spectra were of the same resolution as the comparison stars. 
\subsection{Temperature calibration}\label{sec:T_eff}
\begin{figure}
\begin{center}
\includegraphics[width=0.6\columnwidth]{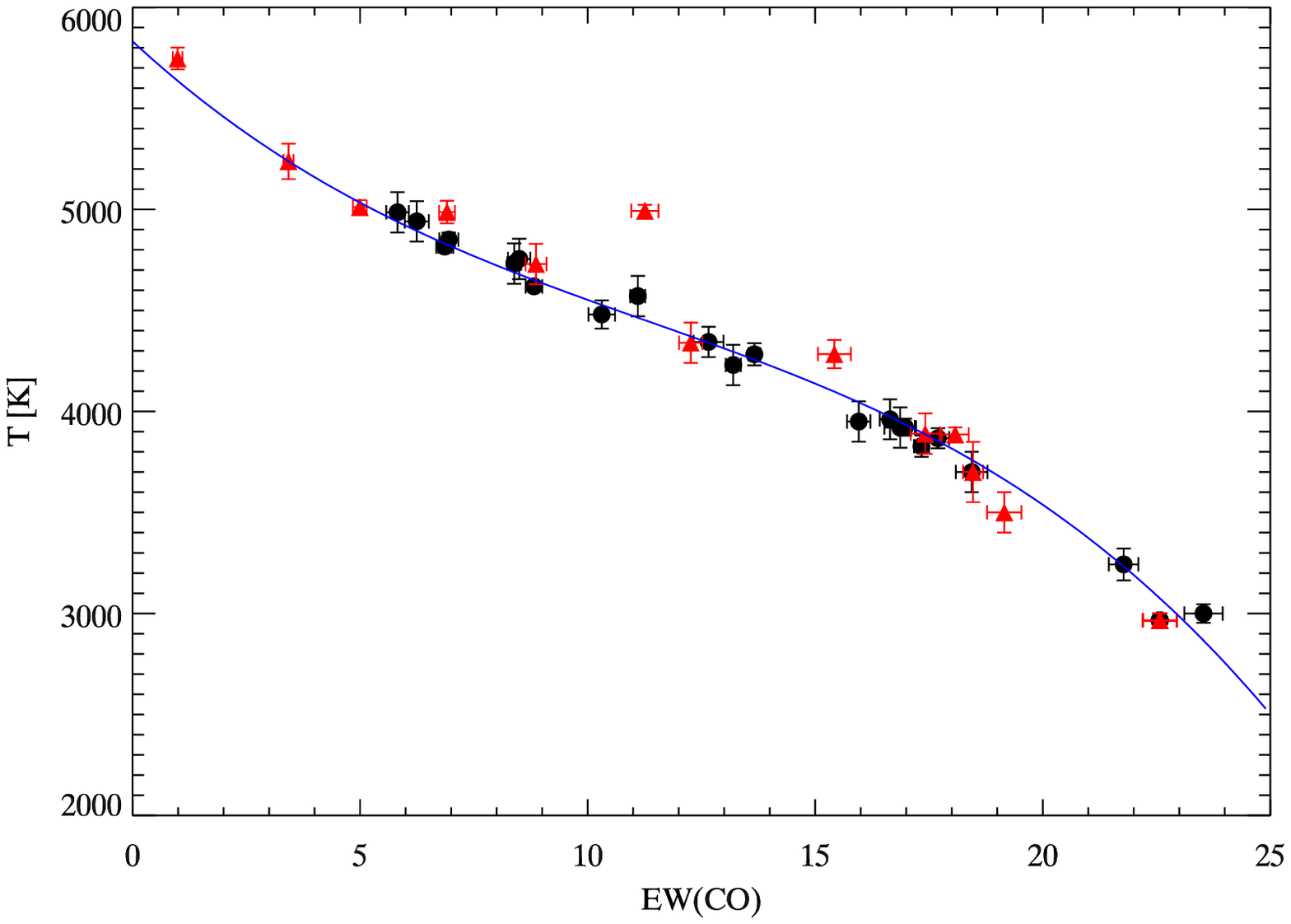}
\caption{The EW(CO)-$\rm T_{eff}$ relation for the template giants is shown. Red triangles indicate spectra with a resolution of $\rm R\sim3000$, black dots represent spectra with a resolution of $\rm R\sim2000$. One $10\,\sigma$ outlier can be seen. This particular star has a metallicity ($\rm [Fe/H]=-0.5$) significantly lower than the other template stars ($\rm <[Fe/H]>=-0.1$). It was only included for comparison, but was not considered in the fit. The solid line shows the best fit to the data. The fit is reliable between $\rm3<EW(CO)< 24$. The fit residual RMS is 119\,K.
}
\label{fig:T_calib}
\end{center}
\end{figure}

For the temperature calibration we used stellar spectra of giants with known \teff. In total we used 33 giants with spectral types G0-M7 and metallicities $\rm-0.3 < [Fe/H] < 0.2 $. The spectra were obtained from the NOAO IR-library \citep{wallace97} with $\rm R\sim3000$; the IRTF-library with $\rm R\sim2000$ \citep{rayner09} and from \cite{for00} with $\rm R \sim2000$. The corresponding temperatures were obtained with the help of the SIMBAD database. 
The CO-\teff~ relation for the template giants can be seen in Fig.\,\ref{fig:T_calib}. The best-fit to the data in the range $\rm 3.5 < CO < 24$ is (where CO is the FR01 index in units of [\AA]): 
\begin{equation}
\teff =5832^{\pm130} - 208.25^{\pm32.84} \cdot {\rm CO} + 11.38^{\pm2.48} \cdot {\rm CO}^2  -0.34^{\pm0.06} \cdot{\rm CO}^3.
\end{equation}
The residual scatter is 119\,K. The average metallicity of the calibration stars of $\rm [Fe/H] = -0.1 $ is somewhat below the average metallicity in the GC of $\rm [Fe/H] =0.14 $ \citep{cunha07}. To account for the metallicity dependence of the CO strength, we add a constant offset of 50\,K to the temperatures of the GC giants (estimated with the help of \cite{marmol08}; their Fig.\,14). 

\subsubsection{Systematic temperature uncertainty}\label{sec:tuncertainty}
\begin{figure}
\begin{center}
\includegraphics[width=0.6\columnwidth]{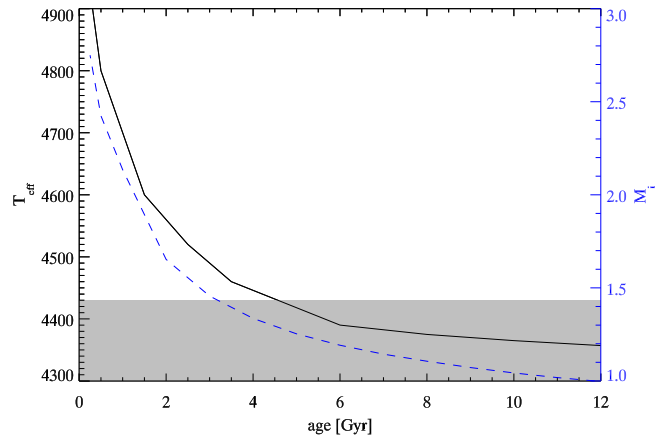}
\caption{Relation between age and median temperature of the Red Clump is shown (solid line). Overplotted is the relation between age and mean initial mass of the Red Clump population (dashed line). The shaded area indicates the \teff~uncertainty. The plot illustrates the temperature sensitivity to the age of the red clump population. For ages $>$5\,Gyrs, the age resolution decreases significantly. 
}
\label{fig:T_eff_RC}
\end{center}
\end{figure}

The systematic errors in the derivation of the temperature are the main drivers in the age uncertainty of a giant population (at a given metallicity). As discussed in the previous section, the CO index computation might be biased on a $\rm \sim 50\,K$ level. More important however is the theoretical uncertainty. The Padua and Geneva Isochrones can differ by up to 80\,K for the same star. Strictly speaking, this is not an observational uncertainty but in the fitting procedure the theoretical uncertainty must be taken into account. We included the bias by assuming a total systematic uncertainty of 100\,K. To visualize the impact of temperature uncertainties, Fig.\,\ref{fig:T_eff_RC} shows the theoretical age\,vs.\,median temperature relation for red clump stars ($ 0 < \mbol < 1$). Especially for old ages, the age interpretation is very sensitive to temperature changes. Small temperature biases of the order 100\,K might change the derived age by several Gyrs. In the star formation history fitting, we used not only red clump stars. However red clump stars make up more than one third of all observed giants. Thus, the systematic uncertainty of the red clump population is representative for the whole giant population.

\subsection{Red clump spectrum: evidence for old ages and near solar metallicity}\label{sec:RC_spectrum}
The majority of the GC stars are red giants (RG). The RG phase is associated with the hydrogen-shell burning phase of a star. While the burning shell propagates outward through the stellar interior, the star expands and cools down. During this evolution, the star increases its brightness and climbs up the RG branch. Discontinuities in the stellar interior cause the brightness of the star to drop temporarily, when the burning shell passes a discontinuity. This happens several times during the ascent of the RG branch. This creates an overdensity of stars at a typical luminosity referred to as the red clump (RC).
\subsubsection{GC red clump}\label{sec:GC_RC}
\begin{figure*}
\begin{center}
\includegraphics[width=0.7\columnwidth]{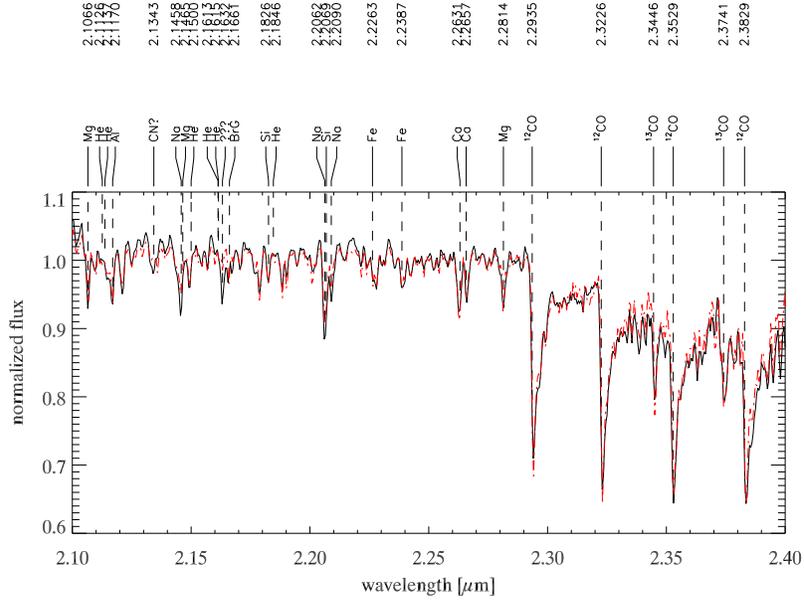}
\caption{The median spectrum of 234 GC red clump giants (black). Each individual spectrum has been normalized and shifted to rest-wavelength. The median spectrum of the GC red clump is closely matched by a K3 giant of the solar neighborhood (red) with a temperature of 4330\,K. The most prominent lines are indicated. Almost every line is actually a blend of atomic and molecular lines. In case of a line blend, the most important line is indicated.
}
\label{fig:RC_spectrum}
\end{center}
\end{figure*}

The observed K-band magnitude and the spectral CO index allowed the determination of the bolometric magnitude and temperature for individual GC stars. The bolometric magnitude \mbol~ follows directly from the intrinsic, de-reddened $M_K$. To derive the intrinsic $M_K$ we used a distance modulus of 14.6 (8.3\,kpc). The extinction was calculated for each star individually, following \cite{bartko10}. We used for each star the 20 nearest neighbors to derive the median H-K colour. By assuming an intrinsic colour of $\Delta m_{H-K}=0.1$ (typical for K giants) we derived the reddening. By assuming a powerlaw extinction $A_{\lambda}\sim \lambda^{-2.21}$ (Sch\"odel et al. 2010, Fritz et al.(2011) we computed the $A_K$ extinction for each star. The bolometric correction ($BC_{K}$) was taken from \cite{blum03} $[BC_{K}=2.6-(\teff-3800)/1500]$. The assumed $BC_{K}$ is consistent with the bolometric correction library from \cite{lejeune97} used in the subsequent population synthesis models. The bolometric $1\,\sigma$ uncertainty is $0.32$ magnitudes. It is caused by the photometric error, the temperature- and the extinction uncertainty. The stellar \teff~ follows directly from the CO strength. The statistical $1\,\sigma$ \teff~ uncertainty derived by error propagation of the measured CO index uncertainty is 200\,K. The systematic \teff ~uncertainty is 100\,K (see Sec.\,\ref{sec:tuncertainty}).
The red clump of the GC giants can be found at a luminosity of $\mbol=0.6$ and a temperature of $\teff=4310\rm \,K$ (see Fig.\,\ref{fig:HRdiag}). Since the red clump is narrow, both in magnitude and temperature, i.e. all stars have very similar spectral features, it is a viable approach to construct a median spectrum of the GC red clump stars. In order to construct a high SNR spectrum, we used 234 red clump stars with bolometric magnitudes $\rm 0 < \mbol < 1$ corresponding to $15.1 < m_K < 16.1$. The average SNR of individual spectra is between 10 and 15. All spectra were normalized and shifted to rest-wavelength as described in the previous section. Most of the spectra have a resolution of $R\approx2000$. Spectra with a higher resolution were interpolated to match the latter. The equivalent integration time of the combined spectrum is of the order 300~hours. Figure\,\ref{fig:RC_spectrum} shows the resulting median spectrum of the GC giants. 

\subsubsection{Temperature estimate from high SNR spectrum}
\begin{figure}
\begin{center}
\includegraphics[width=0.6\columnwidth]{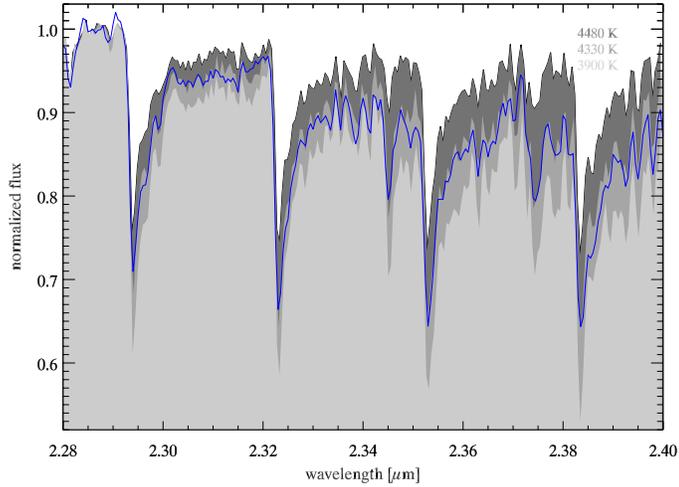}
\caption{The temperature dependence of the CO bandheads is shown. The median spectrum of 234 GC red clump giants (blue) is compared with three giants of the solar neighborhood with temperatures between 4480 and 3900\,K (gray shaded areas). 
}
\label{fig:RC_CO}
\end{center}
\end{figure}

We compared the median spectrum with available library stars. The template star matching the co-added spectrum most closely is a K3\,III giant with a temperature of 4330\,K. The agreement of the spectra is impressive. Only the sharpest peaks are slightly smeared out due to a residual velocity error. The $\rm ^{12}CO$ bandheads at 2.294, 2.322, 2.352 and 2.383\,$\rm\mu m$ are well represented. Even the weaker bandheads of $\rm^{13}CO$ at 2.345 and 2.373\,$\rm\mu m$ match the template (Fig.\,\ref{fig:RC_spectrum}). The comparison with colder and warmer giants (Fig.\,\ref{fig:RC_CO}) shows the sensitivity of the CO bandheads to the stellar temperature. The median temperature of the red clump as derived by the stacked spectrum $\teff=4330\, \rm K$ agrees very well with the median value  ($\teff=4310\, \rm K$) of the individual low SNR spectra. 
The Ca\,I and Na\,I line blends are moderately sensitive to temperature as well as to the surface gravity \logg. The Ca\,I lines are in good agreement with the template. The same is true for the iron line blends at 2.22 and 2.23\,$\rm\mu m$ as well as Mg\,I, Si\,I and Al\,I. Unfortunately almost all of the atomic lines are contaminated with weaker atomic lines and lines of CN. This excludes the possibility of a spectral synthesis fit at the given resolution. However qualitatively, it is clear that the GC spectrum is closely matched by a solar metallicity spectrum, in agreement with the work of \cite{cunha07}. 
\subsubsection{Sodium line blend: Potential evidence for mixing processes}
The strongest deviation between our average spectrum and the K3\,III template is found in the  Na\,I lines. They seem to be intrinsically stronger than in the solar neighborhood. None of the template spectra in the temperature range of the red clump can reproduce the Na\,I strength. An increased Na\,I strength was reported previously in low-resolution spectra by \cite{blum96}. The Na\,I lines are actually blends of a couple of atomic lines and CN lines. The individual contribution depends on the stellar temperature. For temperatures similar to a K3\,III giant, sodium is the most important line. Cooler spectra are heavily influenced by Sc and the coolest spectra are dominated by CN. For a detailed analysis of lines in K-band spectra see \cite{wallace96}. \cite{carr00} used high resolution spectra of IRS\,7 to derive abundances. They found that the increased line strength of the Na\,I complex is mainly caused by stronger CN lines. The CN lines reflect extreme CNO abundances in the atmosphere of IRS\,7, which they claim is probably the result of increased rotational mixing. \\
\cite{cunha07} also find CN-cycled material in the atmospheres of three luminous GC giants, however not as deeply mixed as in IRS\,7. Increased rotational mixing, is predicted for dense stellar clusters, where tidal spin-up can lead to significantly higher rotation speeds of main-sequence stars \citep{alexander05}. This might explain the increased CNO material in the outer atmosphere layers of the luminous GC giants. Evidence for increased rotational mixing taking place in a supergiant like IRS\,7 with a mass of 20\,\msun~ and a lifetime of a few Myrs, are hard to transfer to the red clump giants of the GC with masses of only 1-2\,\msun~ and ages of Gyrs. This is especially true since IRS\,7 is significantly cooler (3600\,K) than the red clump. 
\\
However, \cite{maeder00} note that red giants with masses $M<1.5\,\msun$ are susceptible to extra-mixing and \cite{alexander05} notes that tidal spin-up is most effective in long-lived low mass stars. Thus, the strong sodium lines might indicate that the red clump stars have undergone a mixing process different from the solar neighborhood, for example due to tidal spin-up. In principle, the sodium line strength can also reflect a peculiar chemical composition of the Galactic Center. With the current knowledge we cannot rule out that possibility. However, rotational mixing provides a convincing explanation for the increased sodium strength. This might provide interesting conclusions on the stellar evolution since fast rotating stars tend to be more luminous and redder and can provide evidence for the existence of an underlying dense stellar cusp of low-mass main-sequence stars \citep{alexander05}. 
\\
Apart from the Na\,I lines, the average spectrum shows no peculiarities. Small deviations of the GC spectrum at 2.219 and 2.349\,$\rm \mu m$ are caused by poorly subtracted nebular emission of the Mini-Spiral. The deviation at 2.317\,$\rm \mu m$ on the other hand coincides with a telluric line. Overall the agreement is intriguing and validates the method of constructing a median spectrum a-posteriori. 

\section{Construction of the H-R diagram}\label{sec:const_HRdiag}
\begin{figure*}
\begin{center}
\includegraphics[width=0.48\textwidth]{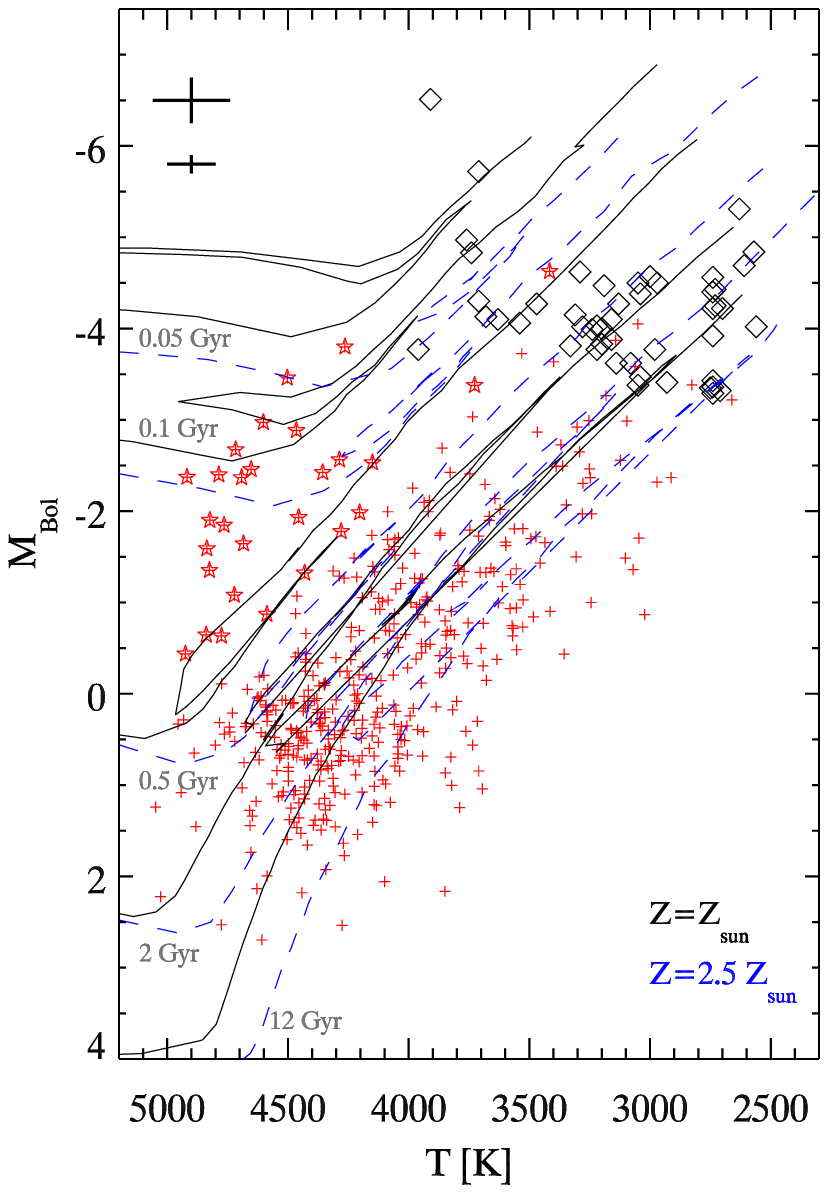}
\includegraphics[width=0.48\textwidth]{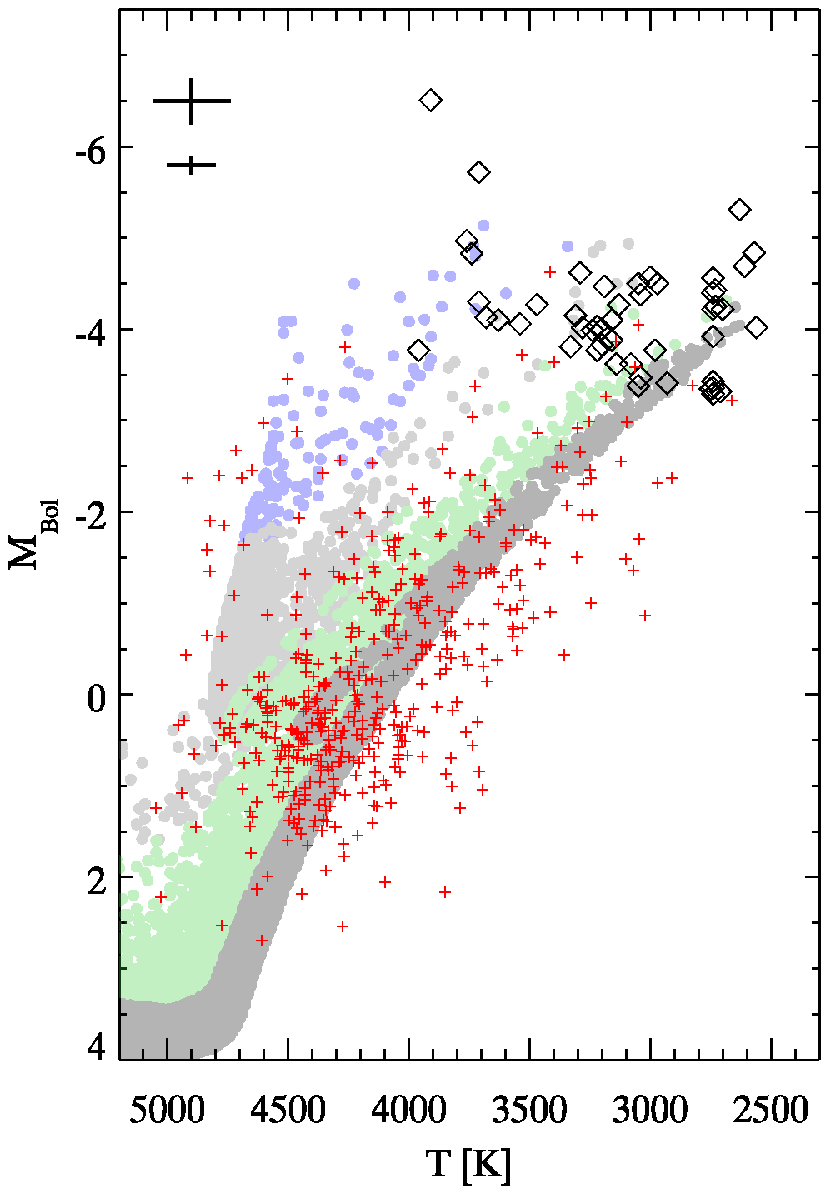}
\caption{{\bf Left:} H-R diagram of the cool Galactic Center stars; this work (red crosses). Stars younger than 0.5\,Gyr are indicated by star symbols; data points from \cite{blum03} are indicated as black diamonds. Overplotted are isochrones for ages (from left to right) 0.05, 0.1, 0.5, 2 and 12\,Gyrs with solar metallicity (black, solid) and 2.5 times solar metallicity (blue, dashed). The upper black cross shows the statistical error and the lower cross shows the systematic uncertainty. {\bf Right:} Observed H-R diagram (red crosses) overlaid on a synthetic model population assuming 12\,Gyrs of continuous star formation; the metallicity is $Z = 1.5\,Z_{\odot}$ \citep{cunha07}; Different age intervals of the synthetic population are indicated; $5<t<12\,\rm Gyrs$ (dark gray); $1<t<5\,\rm Gyrs$ (green); $0.2<t<1\,\rm Gyrs$ (gray); $0.05<t<0.2\,\rm Gyrs$ (blue). The synthetic population was created with the code of \cite{aparicio04}. The code used the \cite{bertelli94} stellar evolution library and the \cite{lejeune97} bolometric correction library.
}
\label{fig:HRdiag}
\end{center}
\end{figure*}

We constructed an H-R diagram for the GC cluster using the same distance modulus and extinction as described in Sec.\,\ref{sec:GC_RC}. The H-R diagram is displayed in Fig.\,\ref{fig:HRdiag}. For comparison, we also included the data of \cite{blum03} probing the luminous giants and supergiants in the inner 2.5\,pc. The comparison data were rescaled to the updated extinction values used in this work. Although \cite{blum03} and this work rely on different temperature calibrations, the overlap regions match very well. The GC population is compared to isochrones \citep{bertelli94} with $Z= Z_\odot$ and $Z=2.5\,Z_\odot$ metallicity (Fig.\,\ref{fig:HRdiag}, left). The right panel of Fig.\,\ref{fig:HRdiag} shows the comparison with with a synthetic model population. The model population represents the case of continuous star formation over 12\,Gyrs with a GC metallicity of $Z=1.5 Z_{\odot}$ \citep{cunha07}. 

\subsection{Features of the H-R diagram}
The H-R diagram shows several distinct features. The most prominent feature is the RG branch consisting of old ($\rm >1\,Gyr$) stars with masses of the order 1\,\msun. The GC red clump can be identified as an overdensity of stars at a luminosity of $\mbol=0.6$. This agrees very well with the red clump in the solar neighborhood measured by \cite{groe08}. The mean red clump temperature derived by individual spectra agrees very well with the estimate based on the high SNR median red clump spectrum (Fig.\,\ref{sec:RC_spectrum}). A second overdensity can be found at $\mbol=-0.8$ and $\teff=3900\,K$. This feature is sometimes referred to as the AGB bump. All afore mentioned features are tracers of an old population. The H-R diagram shows a second branch of giants at $\teff=4800\, \rm K$. This warm giant population is bright ($ M_{\rm bol} < 0$) and separated from the cold (old) branch in the same magnitude range. Yet, only $\sim 10\%$ of the giants between $0 < \mbol <-4$ (corresponding to $11.5 < m_K < 15.5$) can be attributed to the young branch. Similar features are known in globular clusters with multiple stellar populations.  
The comparison of the GC H-R diagram with Chabrier$/$Kroupa populations of different ages can be seen in Fig.\,\ref{fig:HRdiag} (right). Several obvious features are matched. The red clump is well represented in the data. The cool branch of giants with ages greater than 1\,Gyr is obvious. The temperature and luminosity of the warm branch is matched by giants younger than $\rm 0.5\, \rm Gyrs$. A deficiency of stars with ages of $\sim$1\,Gyr can be seen as a gap in the diagram. We don't detect a significant Horizontal Branch of old and metal poor giants as would be typical for old globular clusters. This supports the assumption of a predominantly old and metal rich population. 
\subsubsection{Outliers}
The coolest temperatures we find cannot be reproduced by even the oldest solar metallicity isochrones. Those outliers might be explained by a population of metal rich stars ($Z>2.5\, Z_{\odot}$). However, this does not contradict the current assumption of a near solar metallicity in the GC. The cold outliers account for less than 8\% of the total population. Since metallicity studies in the GC \citep{ramirez00,cunha07} have probed not more than ten late-type stars, a small metal rich population might have gone undetected. This can be compared to the situation in the Bulge (Baade's Window), where a high metallicity tail of stars ($Z>2.5\, Z_{\odot}$) makes up about 15\% of the population \citep{zoccali08}. Another explanation might involve stellar model uncertainties. Most of the outliers show the same luminosity as bright and cool AGB stars. Stellar models of these stars still suffer from large uncertainties. Model isochrones treat this evolutionary phase in a simplified manner and are thus rather unreliable. Furthermore, stars in these stages are known to pulsate with periods of hundreds of days. During the pulsation, an individual star can change its temperature by up to 500\,K \citep{lancon02}. All systematic effects considered by us, lead to an overestimation rather than an underestimation of the stellar temperature. Yet, some of the stars can be heavily dust obscured and thus appear too faint for the given temperature. In the following fitting procedure we ignored stars that were cooler than allowed by the isochrones and the temperature uncertainty. This excludes only a small number of the old stars. Therefore we are confident that the impact on the results is negligible. 
\subsubsection{The young giant branch}
Roughly 10\% of the observed GC giants with magnitudes between $0 < \mbol <-4$ appear to belong to a branch of young ($\rm<500\,Myrs$) giants. With masses between $2.5\,\msun <M < 6\,\msun$ those stars are descendants of main sequence B-stars. They are tracers of an intermediate age population in the Galactic Center.
Their stellar age is smaller than the typical non-resonant two-body relaxation time in the Galactic Center of the order $t_{\rm nr} \gtrsim \rm few~ Gyrs$ in the GC \citep{alexander05,merritt09}. However, close to the SMBH, orbits are near-Keplerian. This causes interactions between stars to build up coherently. The randomization of the angular momentum vectors happens therefore on the fast vector resonant-relaxation timescale $t_{\rm vr} < \rm few ~ten~ Myrs$ \citep{hopman06}. \\ However, the eccentricity and semi-major axis distribution is randomized on the slow scalar resonant timescale $t_{\rm sr}$ \citep{hopman06}, comparable to $t_{\rm nr}$ for $r > 0.1\,{\rm pc}$. 
The mean stellar mass of the young giants ($3.2\,\msun$) is significantly larger than the average mass of the old giants ($1\,\msun$). The more massive stars tend to sink to the center of a cluster due to dynamical friction exerted through the drag of lighter background objects. The mass segregation timescale in the Galactic Center is larger than in normal stellar clusters due to the presence of the SMBH and the corresponding higher velocity dispersion. The mass segregation timescale $t_s\sim t_{\rm nr} \frac{\overline{M}}{M_{\star}}$ scales with the relaxation time $t_{\rm nr}$ but is also a function of the individual stellar mass $M_{\star}$ and the mean stellar mass $\overline{M}$ \citep{alexander05}. Assuming the mean stellar mass is $\overline{M} \approx 1\, M_{\odot}$, the segregation time is $t_{\rm s} \approx \frac{1}{3}t_{nr} > 1\,\rm Gyr$. This is still significantly larger than the age of the young giants. Thus the radial distribution could not have changed significantly since the stars formed (neither through relaxation, nor through mass segregation). Yet, the angular momentum vectors (i.e the orientation of the orbits) will have undergone randomization. 
\subsubsection{Kinematics and distribution}
\begin{figure}
\begin{center}
\includegraphics[width=0.6\columnwidth]{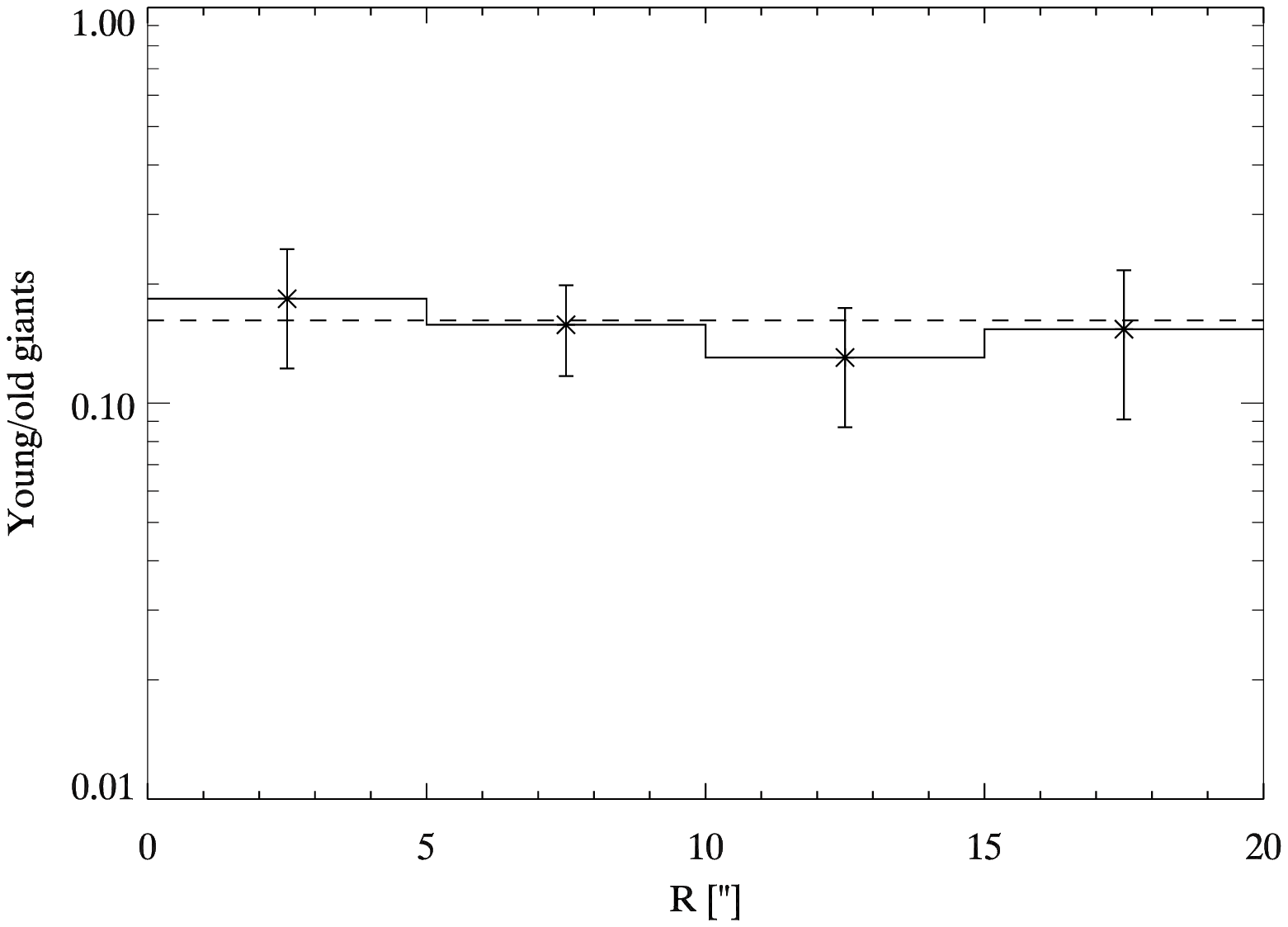}
\caption{Ratio of young-to-old giants with magnitudes between $\rm 0 < \mbol <-4$ as function of radius. The average ratio is indicated (dashed line). The errors are the Poisson errors of the young giants in the given radial bin. The ratio is consistent with being flat, indicating that the young giants exhibit the same radial distribution as the old giants. 
}
\label{fig:rad_dist}
\end{center}
\end{figure}

To assess the dynamical state of the young giants we computed their orbital distribution. We followed the method of \cite{bartko09}, who used statistical arguments to infer the 3D distribution of a stellar population out of the 3D velocity and projected distance distribution. A coherent motion within a stellar system can be detected as a statistical preference for a angular momentum direction. However the orbital distribution of the young giants shows no significant excess as would be the case for a disk or a streamer. The orientation of the angular momentum vectors is consistent with being isotropic, as expected due to the fast vector resonant relaxation. The radial distribution of the young giants compared to the cool giant population is shown in Fig.\,\ref{fig:rad_dist}.

\begin{figure}
\begin{center}
\includegraphics[width=0.6\columnwidth]{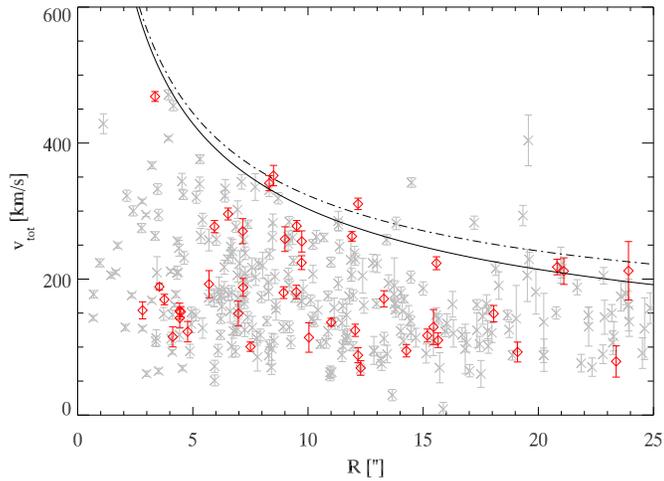}
\caption{The 3D velocity of the warm giants (red) with magnitudes of $\rm 0 < \mbol <-4$ and the cold comparison sample (gray) is shown. The maximum allowed velocity for stars to be bound is indicated; SMBH alone (solid line); SMBH + cluster mass (dashed line). 
}
\label{fig:vel}
\end{center}
\end{figure}

 Both populations exhibit the same radial density distribution. The density distribution of the total giant population has been described as a core or even a hole close to the SMBH \citep{buchholz09,bartko10,schoedel10a}. In any case, the distribution differs significantly from a Bahcall-Wolf cusp, the predicted final state of a relaxed population. Given the age of the young giants, they must have formed rather close to the SMBH, either in-situ or transported in by in-falling clusters. Since the radial density distribution evolves only slowly on timescales of $\rm \sim few \,Gyrs$ this means that the young giants still contain information on their initial distribution. The 3D velocity of the young giants compared with the cool giants in the same magnitude bin is shown in Fig.\,\ref{fig:vel}. The 3D velocity was computed, assuming a distance of 8.3\,kpc to the Galactic Center. The maximum 3D velocity $v_{\rm max} = \sqrt{\frac{2GM_{\bullet}}{r}}$ for stars to be bound is indicated in the figure. The projected distances r yields a lower limit for the physical 3D distance R. The kinematics of the young population are consistent with the old stars. Using a Kolmogorow-Smirnov test to asses the likelyhood that the young giants 3D velocities can be drawn from the cold population returns a probability of 70\%. 
The fact that both giant populations share the same radial distribution is surprising. The young giants ($M_{\rm ZAMS}\approx 3\,\msun$) are three times more massive than their old counterparts with a lifetime too short to experience relaxation and especially to change their angular momentum significantly. However, it renders a radially varying IMF unlikely. The latter is a valid claim although star formation history and IMF are highly degenerate, since one effect would need to cancel exactly the other to emulate a radially constant distribution of young-vs-old giants.

\section{Calculation of the star formation history}\label{sec:SFR_fit}
To derive the star formation history, we compared the observed distribution of giants in the H-R diagram to model populations. We restrict our analysis to the $m_K<15$ magnitude range with $>50\%$ completeness. This left us with 450 giants for the fit. We repeated the calculation with a completeness limit of 80\%. The results were similar, yet with larger uncertainties. Thus we are confident that no bias is introduced due to the completeness limit chosen. The synthetic model populations were created with the IAC-STAR code \citep{aparicio04}. The code allows a selection of different stellar evolution and bolometric correction libraries for the computation. We chose the \cite{bertelli94} stellar evolution library and the \cite{lejeune97} bolometric correction library since these are the only libraries with evolutionary tracks of stars with masses $>10\,\msun$. The code requires several input parameters such as metallicity, age and slope of the IMF. Since age and metallicity introduce some degeneracy, we adopted the mean metallicity found by \cite{cunha07}. Thus, we set up model populations with a metallicity of $ Z = 1.5 \,Z_{\odot}$. 
For the calculation of the star formation history, we applied a method similar to \cite{blum03}. We defined four age bins (50-200\,Myrs, 0.2-1\,Gyr, 1-5\,Gyrs and 5-12\,Gyrs) with constant star formation rate within each age bin. The bins were chosen such that the evolutionary tracks are distinct enough to be resolved with the given data quality. The isochrones for stars with ages $\rm >few \,Gyrs$ are very similar. Therefore, the older age bins were chosen to be wider than the younger age bins. The younger age bins were selected by visual comparison of the model populations with the data. Fig.\,\ref{fig:HRdiag} shows the chosen age bins. To study the impact of the IMF, we created models with various IMF slopes between $-2.7\le \alpha \le-0.45$. Among the models, we included a Chabrier$/$Kroupa IMF ($\alpha=-2.3$), a top-heavy IMF ($\alpha=-0.85$) \citep{maness07} and a flat IMF with a slope of $\alpha=-0.45$ as it was found recently for the young stellar disk \citep{bartko10}. We note, that the star formation could have proceeded in episodic bursts. Yet, given the quality of the data, it is not possible to distinguish between a burst and a continuous formation within an age bin. Therefore the derived star formation rate is an average across the age bin.
\subsection{Fitting procedure}\label{sec:fit}
We added Gaussian noise to the synthetic model populations, representing the statistical errors of \mbol~ and \teff. Then we removed stars from the synthetic populations according to the estimated completeness as a function of K-band magnitude. The completeness is not a function of temperature, since we used a minimum SNR criterion. We then binned the data and the model populations into an H-R diagram with a binsize of $1.5\,\sigma$ of the typical errors. The data were then fit as a linear combination of the H-R diagrams of the 4 model populations (4 age bins). For the fit we used the IDL routine TNMIN \citep{markw08}. As a minimization parameter we used the Poisson maximum likelyhood parameter $\chi^2_{\lambda}=2\sum_i m_i-n_i+n_i ~\rm{ln}(n_i/m_i)$, where $m_i$ is the number of stars predicted by the model and $n_i$ is the number of observed stars in the $i$th bin of the H-R diagram \citep{mighell99,dolphin02}. The contribution of each age bin determines the relative star formation rate of that bin. The best-fitting star formation rate for various IMFs is given in Table\,\ref{tab:SFR}. 
\subsubsection{Uncertainty and quality of the fit}
\begin{figure}
\begin{center}
\includegraphics[width=0.9\columnwidth]{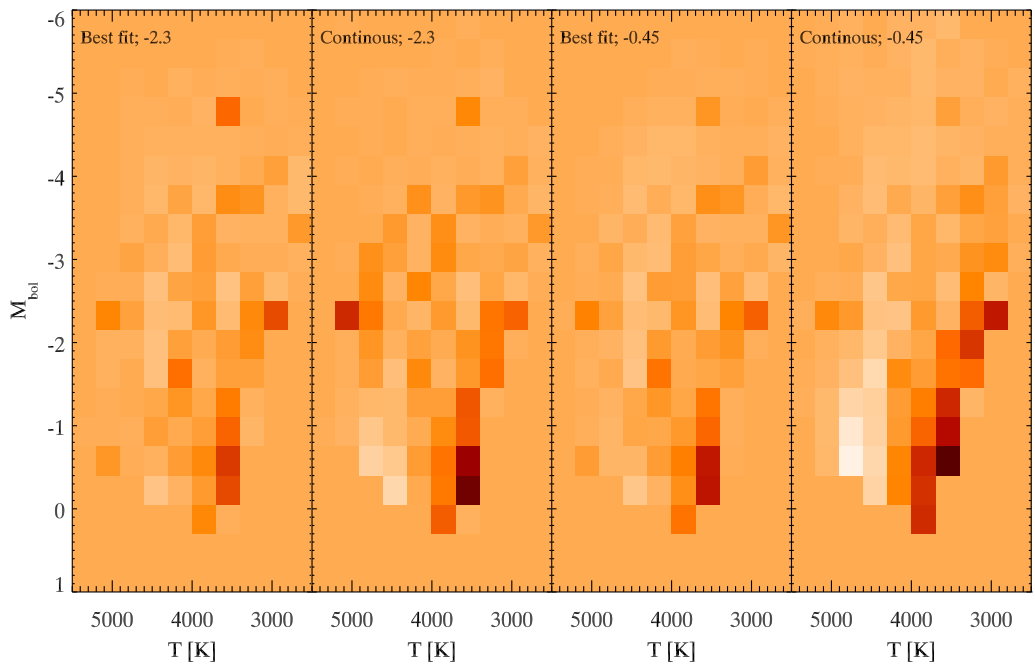}

\caption{Residual diagrams of two best-fit models (with a normal Kroupa IMF, $\alpha=-2.3$ and a flat IMF, $\alpha=-0.45$) compared with models of continuous star formation and the same IMFs. Each panel shows the residual model - data H-R diagram weighted by the variance in each H-R bin. Lighter areas indicate the model overpredicting the number of stars in the corresponding bin compared to the actual data. Darker regions indicate bins where the model predicts too few stars compared to the data. It is apparent that the data is well matched by models allowing for a time varying star formation rate. The continuous star formation models predict too few old stars on the cold giant branch, while they overpredict the number of young and massive giants on the warm branch. The discrepancy is worse in case of the flat IMF. The best-fit models however reproduce both branches and show only little discrepancies.}
\label{fig:hess_diagram}
\end{center}
\end{figure}

To assess the fit uncertainty we used a method generally referred to as bootstrapping. We constructed 1000 H-R diagrams by drawing random stars out of the data. Each star was allowed to be drawn any number of times. To include the systematic uncertainty of the data and the theoretical isochrones, we added a random temperature offset with a Gaussian $\rm \sigma$ of 100\,K to the whole data set. The constructed H-R diagrams were fit again with the model populations. The scatter of the derived relative star formation rates represents the $1\,\sigma$ uncertainty of the relative star formation rate (see Table\,\ref{tab:SFR}). Note, the star formation rate uncertainty is mainly driven by the systematic temperature uncertainty. To judge the quality of each model fit, we produced 1000 Monte Carlo data sets, each containing 450 stars drawn randomly from the respective model (not the original data). Calculating the variance in each bin resulted in a variance diagram. We defined the fit quality as $\chi^2=\sum_i \frac{(m_i-n_i)^2}{\sigma_i^2}$, where $\sigma_i^2$ is the variance in bin i, $m_i$ is again the number of model stars and $n_i$ the observed number of stars in bin i. The values for $\chi^2$, the number of degrees of freedom and the corresponding probability that the data can be drawn from the model population are represented in Table\,\ref{tab:SFR}. The numbers of degrees of freedom are equal to the number of populated bins minus the number of free parameters. Fig.\,\ref{fig:hess_diagram} illustrates the residuals for various models. Each panel shows the residual model - data weighted by the variance \citep{dolphin02}. The color coding indicates bins where the model predicts more stars than observed (bright) respectively where the model underpredicts the number of stars (dark). The first panel of Fig.\,\ref{fig:hess_diagram} shows the best-fit model, while the second panel shows a continuous star formation (both assume a Chabrier IMF). The continuous model predicts too few old (i.e. cold) stars compared to the observations, while it predicts too many young (i.e. warm) stars. The data is well fit by a mostly old ($\rm>5\,Gyrs$) population with an admixture of recently formed stars ($\rm<few~ 100\,Myrs$) as in case of the best-fit model. The discrepancy between a continuous formation scenario and a time varying star formation increases in case of IMFs favoring the formation of massive stars (right two panels). 
\section{Results}\label{sec:results}
The data fit best a normal Kroupa$/$Chabrier IMF (49\% acceptance probability, see Tab.\,\ref{tab:SFR}) but either flatter or steeper IMF can be accommodated within the 2 sigma uncertainties. The distribution in the H-R diagram does not strongly constrain the IMF, in contrast to the diffuse light and dynamical mass discussed in Sections \ref{sec:dyn_mass} and \ref{sec:background}.
This is not surprising, since the old giant branch contains stars with zero age main sequence (ZAMS) masses between $1\,\msun < M < 1.2\,\msun$. Thus the mass interval is too small to be significantly affected by changes of the IMF slope. The slope can affect the abundance of young giants, since they cover the mass range $2.5\,\msun < M < 6\,\msun$. However their abundance depends sensitively on the star formation rate of the last few hundred Myrs. Furthermore, the bright and massive end of the population suffers from low number statistics. This makes statements on the IMF slope uncertain. In general, the IMF and the star formation rate as a function of time are largely degenerate. Given the coarse sampling of the age bins, the data is not sufficient to constrain the IMF by the fitting procedure itself. The fit only allows to derive relative star formation rates for each assumed IMF. However, it is possible to constrain the IMF using the measured dynamical mass. Since each model predicts a certain mass composition of the population, it is possible to distinguish between different models. In any case, no acceptable fit was achieved for continuous star formation scenarios, irrespective of the assumed IMF.
\begin{center} 
\begin{table*} 
\begin{center} 
\caption{The star formation rate for each age bin and various model IMFs. \label{tab:SFR}} 
\begin{tabular}{lcccccc} 
\hline\hline 
Model  & Age bin [Gyr] & IMF slope \tablenotemark{a}  & $ \chi^2/\rm d.o.f.~(prob.)\tablenotemark{b} $& Absolute SFR $[10^{-4}\msun yr^{-1}]$ & $\sigma_{\rm SFR}$ \\ 

\hline    
Model 1  & 12 - 5 & -2.7 (-1.3)   & 122$/$94 (3\%)  & 3 & 0.6  \\
 & 5 - 1 & ...   &  ...  & 0.3 & 0.5  \\
 & 1 - 0.2 & ...   &  ...  &  1 & 0.5  \\
 & 0.2 - 0.05 & ...   &  ... &  12 & 3  \\
Model 2 (Chabrier) & 12 - 5 &-2.3 (-1.3)    & 105$/$104 (49\%)  & 3 & 0.6  \\
 & 5 - 1 &...   & ... & 1 & 0.5  \\
 & 1 - 0.2 &...     & ... & 0.5 & 0.5  \\
 & 0.2 - 0.05 &...     & ... & 8 & 2  \\
Model 3 & 12 - 5 & -1.5     & 157$/$106 (0.1\%)  & 13 & 3  \\
 & 5 - 1 & ...    & ...  & 2 & 1  \\
 & 1 - 0.2 & ...    & ... & 1 & 0.7  \\
 & 0.2 - 0.05 & ...    & ...  & 11 & 3  \\
Model 4 (Top-heavy) & 12 - 5 & -0.85 & 143$/$100 (0.3\%) & 126 & 27  \\
 & 5 - 1 & ...   &  ...  & 24 & 11  \\
 & 1 - 0.2 & ...   &  ...  & 6 & 3  \\
 & 0.2 - 0.05 & ...   &  ...   & 50 & 12  \\
Model 5 (Flat) & 12 - 5 & -0.45    & 131$/$107 (6\%)  & 614 & 124  \\
 & 5 - 1 & ...   &  ...  & 97 & 43  \\
 & 1 - 0.2 & ...   &  ... &  22 & 10  \\
 & 0.2 - 0.05 & ...   &  ... &  145 & 34  \\
Continuous & 12 - 0.05 & -2.3 (-1.3)   & 178$/$105 ($<0.01\%$) & 2 & ...  \\
Continuous & 12 - 0.05 & -1.5    & 181$/$111 ($<0.01\%$) & 6 & ...  \\
Continuous & 12 - 0.05 & -0.85    & 176$/$116 ($<0.01\%$) & 34 & ...  \\
Continuous & 12 - 0.05 & -0.45   & 211$/$115 ($<0.01\%$) & 113 & ...  \\
\hline 
\end{tabular} 
\tablenotetext{a}{The upper and lower mass cutoff is: $ 0.5 < M < 120$. The models 1 and 2 extend to $ 0.1 < M < 120$ with a flatter slope between  $ 0.1 < M < 0.5$. }
\tablenotetext{b}{Outlier corrected $\chi^2$ and degrees of freedom. Outliers are bins off by more than $10\,\sigma$. The d.o.f. corresponds to the number of populated H-R diagram bins (ignoring significant outliers) minus the number of fit parameters. Model 1-5 have four independent fit parameters (age bins), while the continuous models have only one free scaling parameter. In brackets, the acceptance probability is stated.}
\end{center} 
\end{table*} 
\end{center}

\subsection{Star formation rate over cosmic time}\label{sec:sfr_with_time}
\begin{figure}
\begin{center}
\includegraphics[width=0.48\textwidth]{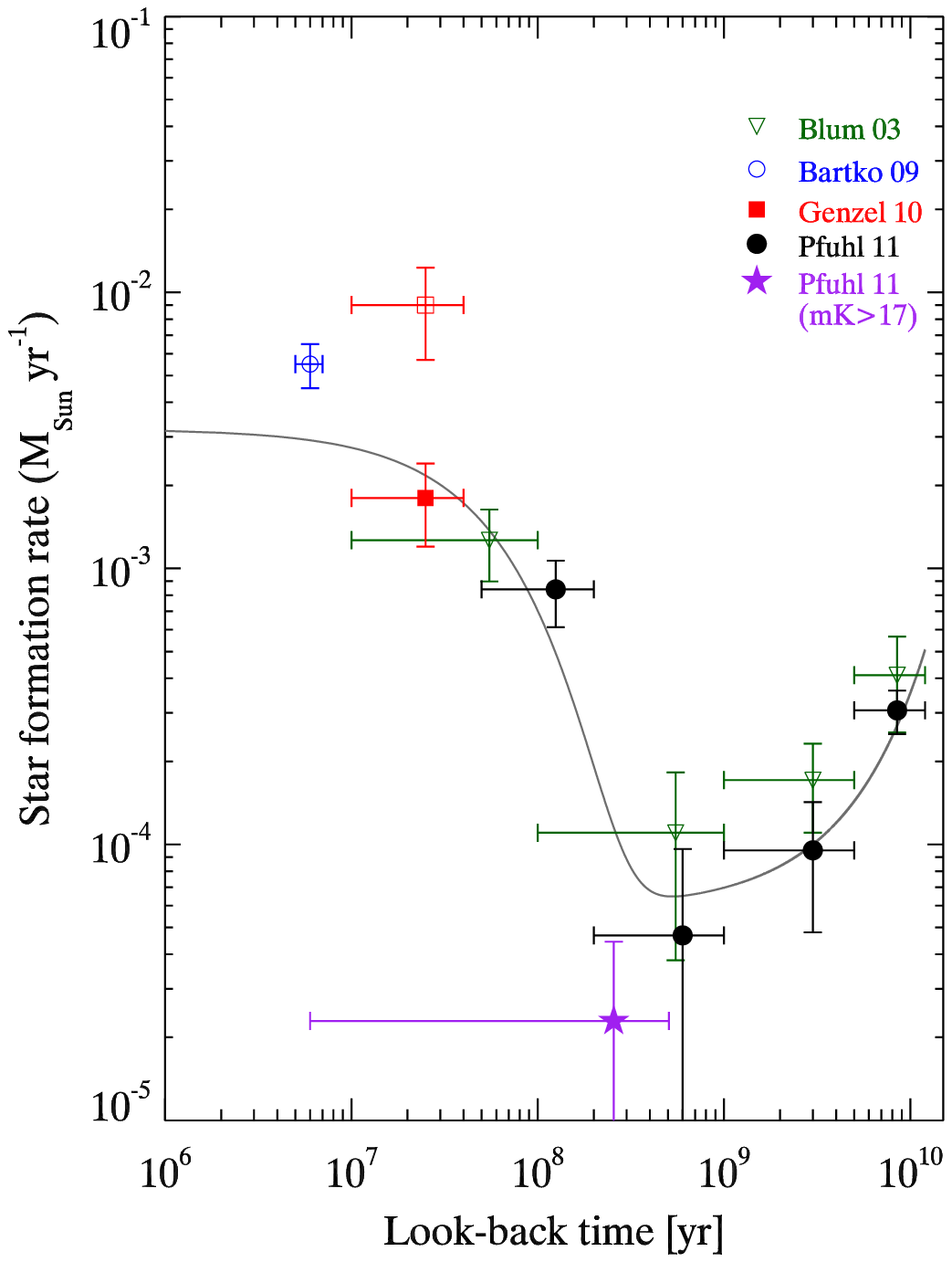}
\includegraphics[width=0.48\textwidth]{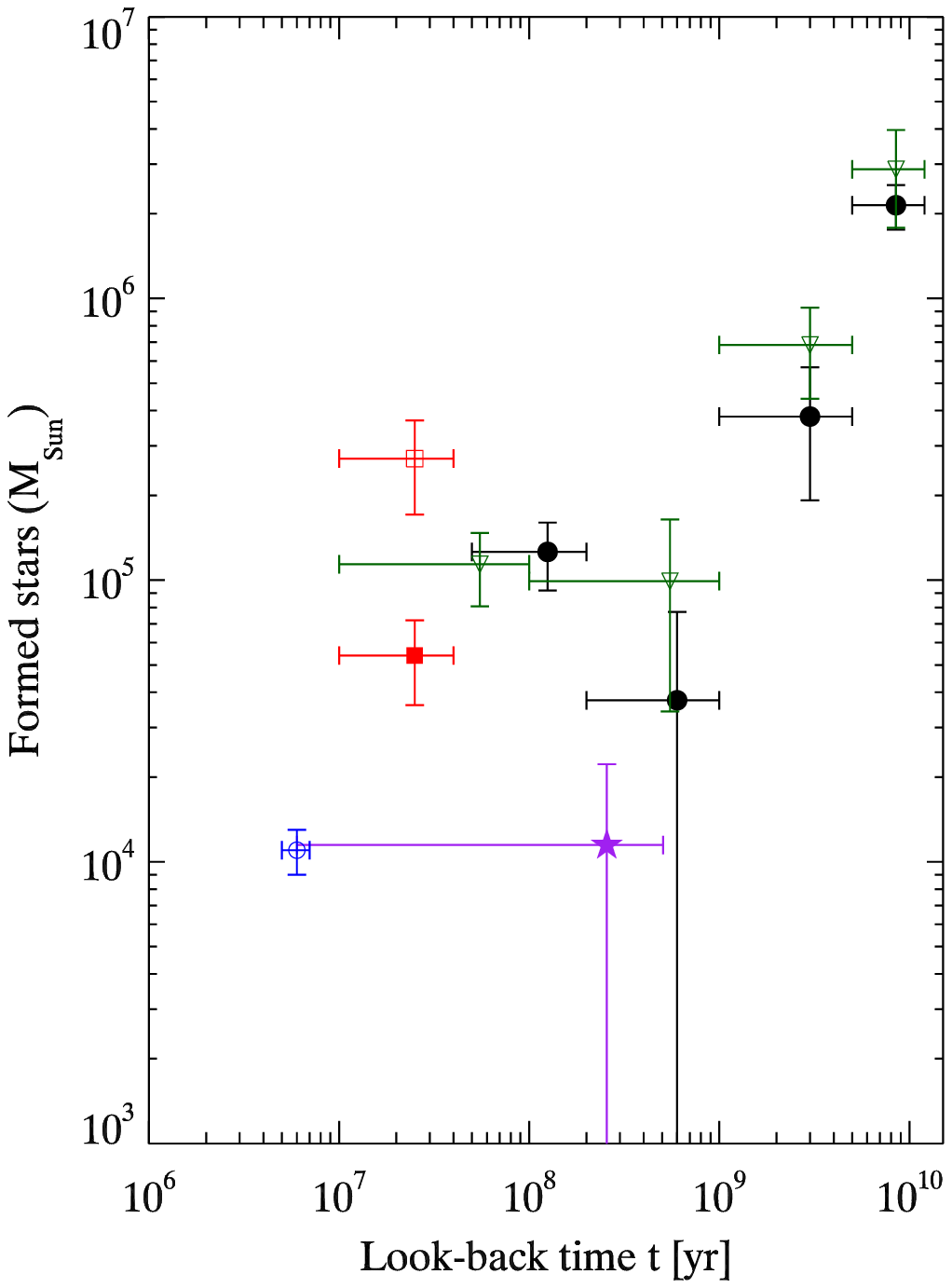}
\caption{{\bf Left:} Star formation rate of the Galactic Center as function of time. The black circles represent the best-fit to the H-R diagram with a Chabrier$/$Kroupa IMF. The mass error is given by the 1$\sigma$ error of the fit. The age error is simply the width of the age bin. For comparison, the star formation history derived by \cite{blum03} scaled to a radius of 1.2\,pc is indicated (green triangles). The star formation rate derived by red supergiants \citep{genzel10} within 1\,pc (red square, filled) and within 2.5\,pc (red square) and the stellar disks (blue circle) \citep{bartko09} are indicated. The star symbol (purple) indicates the recent star formation rate inferred by the early-to-late ratio of faint GC stars (Sec.\,\ref{sec:impl_main}). The solid gray line shows a simple exponential model (see text) of the star formation as function of time. {\bf Right:} The total mass formed in each age bin is shown. Although star formation occured at a high rate during the last few hundred Myrs, the total mass contribution is $<10\%$. The bulk of the stellar mass formed more than 5\,Gyrs ago. 
}
\label{fig:SFR}
\end{center}
\end{figure}

\begin{figure}[h]
\begin{center}
\includegraphics[width=0.4\columnwidth]{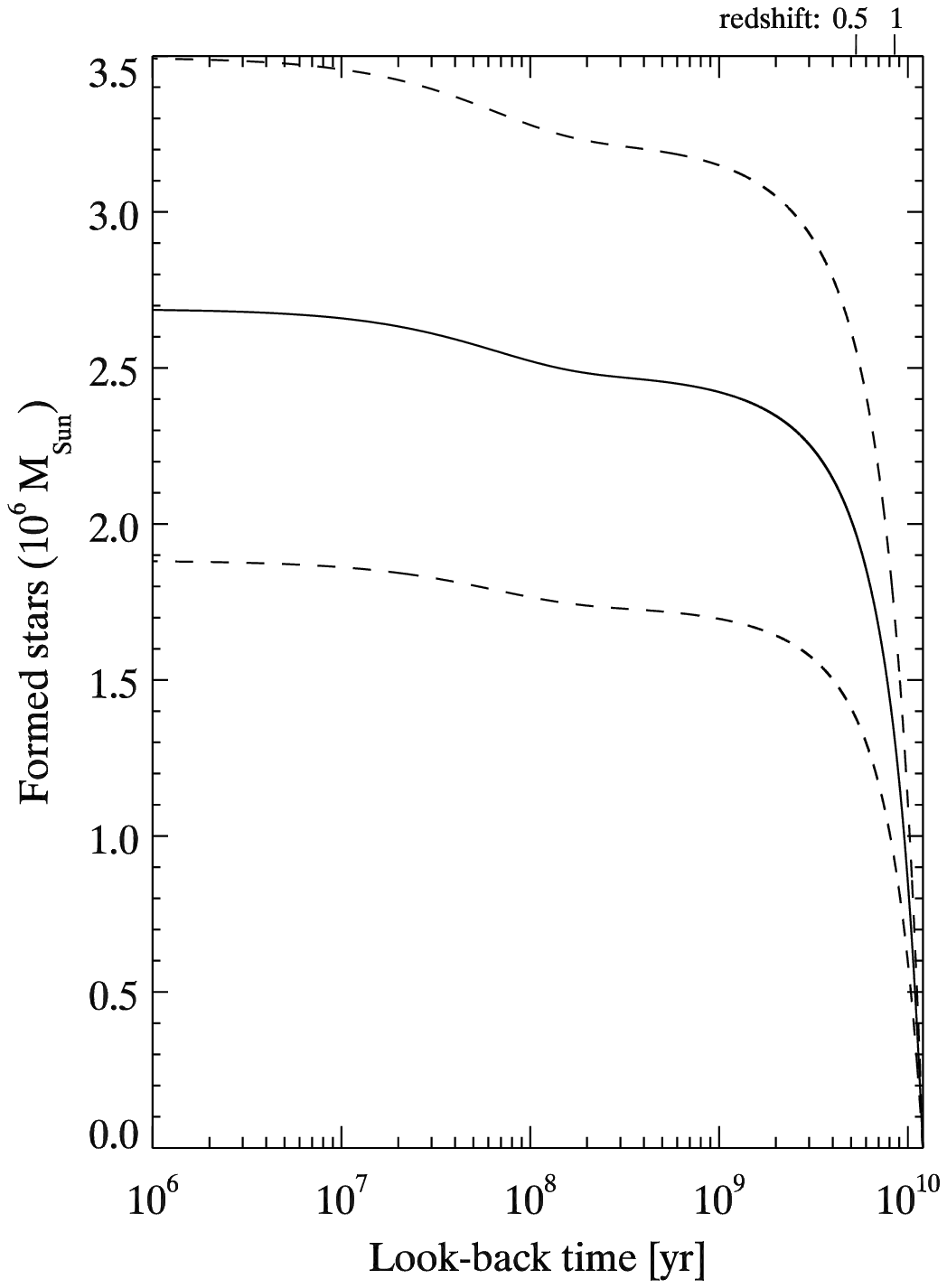}
\caption{The integrated mass as function of time as obtained by Eq.\,3 in Sec.\,\ref{sec:sfr_with_time} is shown. The dashed curves represent the $1\,\sigma$ uncertainty. Note, today's cluster mass is only about 50\% of the total formed mass. The remaining mass is lost due to stellar evolution. 
}
\label{fig:intSFR}
\end{center}
\end{figure}

In the following we only consider the formation scenario with a Chabrier$/$Kroupa IMF. This is motivated by the slight preference of the fitting, the intriguing total mass prediction of the model matching the dynamical measurements and the agreement with the observed diffuse background light (see next sections).
We find that the giant population of the Galactic Center is old. Fig.\,\ref{fig:SFR} shows the star formation rate as function of time (assuming a normal IMF). The formation rate more than 5\,Gyrs ago was on average $3\pm0.6\times10^{-4}\msun/yr$.  According to our fit, roughly 80\% of the total mass formed more than 5\,Gyrs ago. The formation period was followed by a period of reduced star formation lasting another 4-5\,Gyrs. The star formation rate reached a minimum about 1\,Gyr ago. During the last few hundred Myrs, the formation rate increased again. The disk of young stars are part of the increased formation rate. Our results are in excellent agreement with the earlier findings of \cite{blum03} (compare Fig\,\ref{fig:SFR}). The recent period of star formation accounts for about 10\% of the total formed mass. Note, the recent star formation rate seems to be higher than several Gyrs ago. However the bin widths are very different. Each bin represents an average formation rate. The actual formation several Gyrs ago can have happened on short bursts with significantly higher star formation rates. The present day cluster contains roughly 50\% of the total processed gas mass in living stars and 10\% in remnants. The rest has been lost via stellar winds, explosions and potentially been swallowed by the SMBH. For simplicity, we set up a simple analytic model that approximates the measured SFR as function of time (solid line in Fig.\,\ref{fig:SFR}). 
\begin{equation}
{\rm SFR}(t)=6.8\times10^{-5} {\rm \,\msun/yr} \times e^{t/5.5 \,{\rm Gyrs}} + 4.3\times10^{-3}{\rm \,\msun/yr }\times e^{-t/0.06\,{\rm Gyrs}}.
\end{equation} 
where $t$ is the look-back time. The integration of the model yields the integrated mass as function of time.
As Fig.\,\ref{fig:intSFR} shows, about half of the cluster mass formed before a redshift of one. This suggests, that the nuclear cluster formed at a time, when the galaxies build up most of their stellar mass. In that sense, the nuclear cluster formed at the same time, or shortly after the bulge, about 10\,Gyrs ago \citep{zoccali08}.  

\subsubsection{Implications from the faint main-sequence population}\label{sec:impl_main}
The two confirmed A-stars and the one candidate (Sec.\,\ref{sec:Astar_detection}) are ideal tracers for star formation during the last few hundred Myrs. Early-type stars with K-band magnitudes $17<m_K<18$ (assuming $A_K=2.8$ and $ R_0=8.3\,\rm kpc$) are main sequence B9$/$A0 dwarfs with average lifetimes of 500\,Myrs. Late-type stars in the same magnitude range can only be giants that have already left the main sequence. These giants have ages between 1 to 12\,Gyrs. Thus the number count ratio of the two populations provides a measure of the star formation efficiency during the last 500\,Myrs relative to earlier star formation. This ratio is largely independent of systematic uncertainties since it avoids issues like incompleteness and spectroscopic detectability. We included the candidate star, in the early-type population. In total we found three (two confirmed) early-type stars and 30 late-type stars in this magnitude bin. We used the IAC-STAR code and found that in case of continuous star formation over 12\,Gyrs with a Kroupa IMF, the predicted early-to-late ratio is close to unity. The observed ratio therefore argues for a average star formation rate  during the last 500\,Myrs of only about $10\pm 6\%$ of the average formation rate 1-12\,Gyrs ago. To convert the relative formation rate into an absolute rate, we used the star formation rate inferred in the previous section and calculated an average rate of $2.3\pm0.5\times10^{-4}\,\msun$ between 1-12\,Gyrs look-back time. Using this value, we obtained an average star formation rate of $2.3\pm2.1\times10^{-5}\,\msun$ during the last 500\,Myrs (see Fig.\,\ref{fig:SFR}). Within the errors this is consistent with the one derived by the H-R fitting. The A-star ratio provides an independent measurement of the relative star formation rates. It is another indicator for the early formation of the nuclear cluster. 
\subsubsection{Comparison with previous work}
 \cite{maness07} found that the giant population of the GC is on average warm and thus young. Consequently, they favored models with a normal IMF and an increasing star formation rate or a top-heavy IMF and continuous formation. \cite{maness07} used a CO index that is very sensitive to systematic effects. In particular, their CO index, though widely used, was recently discovered to vary with spectral resolution \citep{marmol08}. Since no suitable library was publicly available at the time of writing, they also used template libraries with resolutions between $R\sim3000-5000$ to calibrate their $R\sim2000$ data. The sum of the effects caused an underestimation of the CO equivalent width and an according overestimation of the stellar temperatures. This led to the interpretation of a mostly young giant population. Applying the same methods as \cite{maness07} together with a CO index that is less susceptible to systematic effects and using the newly available library of \cite{rayner09} with $R\sim2000$, allowed us to revisit the star formation history of the Galactic Center.
\subsection{The mass composition of the nuclear cluster}
The present-day mass composition of the nuclear star cluster depends sensitively on the formation history and the IMF. The mass contribution of stellar remnants and stars as well as the total amount of consumed gas depends on the age and the IMF of a population. The consumed gas mass is larger than the sum of remnant and stellar mass because a significant mass fraction is lost due to stellar evolution. Since massive stars lose a larger fraction of their initial mass, the gas consumption increases with flatter IMFs for the same total stellar mass. The same is true for the remnant mass. Massive stars have main-sequence lifetimes  that are significantly shorter than the age of the galaxy. Therefore many generations of massive stars evolve through time and finally end as remnants. Thus, IMFs that favor massive stars produce more and heavier remnants. As a consequence, the mass contribution of stars still burning hydrogen drops with flatter IMF slopes. 
\subsubsection{Constraints from the dynamical mass}\label{sec:dyn_mass}
\begin{figure}
\begin{center}
\includegraphics[width=0.7\columnwidth]{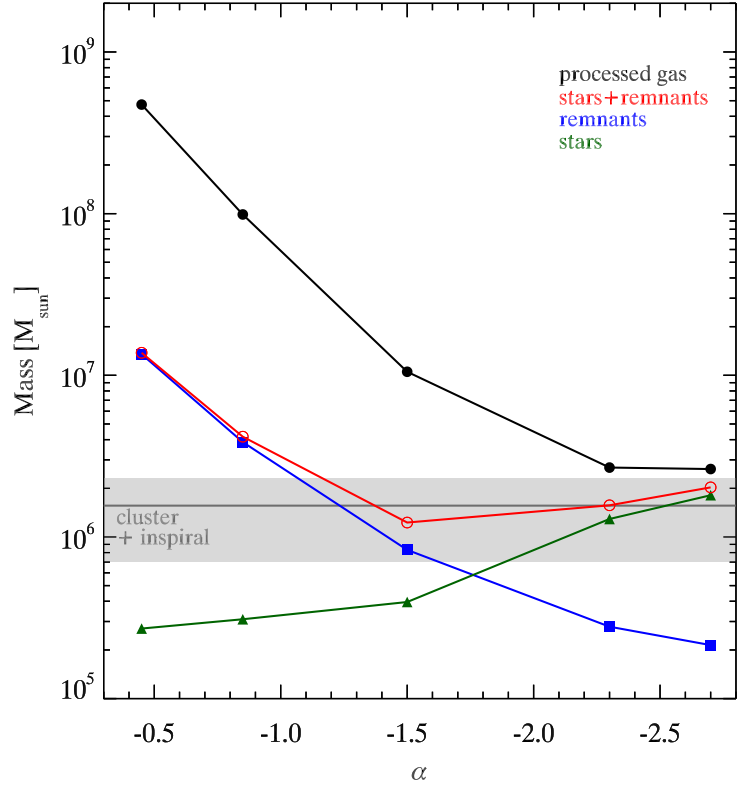}

\caption{Mass composition of the inner 1.2\,pc after 12\,Gyrs of star formation. The composition and total mass depends sensitively on the assumed IMF. The symbols show the five fitted models (Tab.\,\ref{tab:SFR}) with various IMF slopes $-2.7\le \alpha \le-0.45$. Green triangles represent the stellar mass, blue squares represent the remnant mass, red open circles show the sum of remnant and stellar mass and black filled circles show the total processed gas mass. The dark gray horizontal line indicates the dynamical mass within 30\,\arcsec and its uncertainty (light gray area).}
\label{fig:mass_fraction}
\end{center}
\end{figure}

Various models have been fit to the data (Sec.\,\ref{sec:fit}). Each model yields the star formation rate as function of time under the assumption of a certain IMF. To derive the absolute mass contribution, it is necessary to scale the models according to the actual star counts. To get a representative number for the nuclear cluster we used the stellar surface density from \cite{schoedel07}. As discussed in Sec.\,\ref{sec:introduction} the radial extent of the NSC is $\sim 5\, \rm pc$. Our data, however, probed only the inner 30\arcsec~(1.2\,pc). Therefore we restricted ourselves to that radius. By assuming 30\arcsec~as a  sharp edge, the surface density from Sch\"odel et al.\,(2007, their fig.\,12) yields $\sim 18200$ stars with $\rm m_K < 17.75$ within a projected radius of 30\arcsec~ from Sgr\,A*. Between 45\% to 55\% of the projected stars are also contained within a 3D distance of 30\arcsec, depending on the radial density profile derived by \cite{schoedel07}. Thus, we assumed a total of 9100 stars with $\rm R_{3D}<1.2\,pc$ to scale the remnant-, stellar- and total gas mass for each model. Table\,\ref{tab:SFR} shows the derived star formation rate for each model IMF. Fig.\,\ref{fig:mass_fraction} shows the mass composition for the models derived in Sec.\,\ref{sec:fit}. The model predictions have to be compared with the dynamical mass estimates of the nuclear cluster (see Fig.\,\ref{fig:mass_fraction}). The total mass enclosed within 30\arcsec~ is $5.7 \pm 0.9 \times10^6\,\msun$. The SMBH contributes $4.3 \pm 0.5 \times10^6\,\msun$ \citep{ghez08,gill09}, while the remaining $1.4 \pm 0.7 \times10^6 \,\msun$ \citep{genzel96,trippe08,schoedel09,genzel10} are contained in stars and stellar remnants. Numerical simulations of \cite{freitag06} have shown that during 10\,Gyrs about $4\times10^5\,\msun$ stellar and $1\times10^5\,\msun$ remnant mass is removed from the cluster by the SMBH through tidal disruption and inspiral. The cluster composition is scaled to match the observed number of giants. Thus, the inferred stellar mass intrinsically accounts for stars lost to the SMBH because giants and giant progenitors are equally likely to be disrupted as unresolved main-sequence stars constituting the bulk of stellar mass. The situation for remnants, however, is different. The main remnant contribution comes from stellar BH dominating the mass density in the inner 0.1\,pc. Due to their mass $(\sim10\msun)$ they are significantly affected by mass segregation and more likely to inspiral into the SMBH. Consequently, the inspiral of stellar BHs has to be taken into account in the mass budget. We simply added $10^5\,\msun$ with an uncertainty factor two to the observed dynamical mass to constrain the IMF. We took the value determined by \cite{freitag06} as being constant with IMF. This might be the weak point in the line of argument, since \cite{freitag06} assumed a Chabrier$/$Kroupa IMF in their simulations. As Fig.\,\ref{fig:mass_fraction} shows, the cluster composition depends sensitively on the assumed IMF. For flatter IMFs, the number of stellar BH increases, however, the mean mass increases accordingly making mass segregation less efficient. Thus the actual mass transfer between the cluster and the SMBH as function of IMF can only be addressed by further simulations. Keeping the limitations in mind, the dynamical mass plus the removed mass therefore make IMFs with a slope flatter than $\alpha=-1.1$ unlikely. Even if the transferred mass is significantly higher, models flatter than $\alpha=-0.8$ violate the total enclosed mass (SMBH+cluster). Mechanisms that expel remnants or stars are very inefficient. One mechanism is the evaporation of stars from the nuclear cluster. However, for the GC the timescale is greater than a Hubble time \citep{alexander05}. Mass segregation is another mechanism for removing stars. This mechanism moves low-mass constituents outward, and high-mass constituents inward, without changing the radial density profile significantly. \cite{miralda00} suggest that inward migration of a considerable number of bulge stellar blackholes can remove low mass stars from the nuclear cluster. According to their model, the low mass stars are supposed to form a core with a radius of 1-2\,pc. Observations however find a core radius of about 0.3\,pc \citep{schoedel07}. If the observed core is related to mass segregation, then the effect is smaller than predicted.
Our constrains on the IMF are however not weakened by potential mass segregation. Our models use the number of observed stars as a scaling for the total processed mass (and remnants produced). 
Therefore if low mass stars are removed from the inner region, our scaling underestimates the remnant mass. 
The inferred stellar mass however is not changed, because giants and unresolved main sequence stars have very similar masses ($\sim 1\,\msun$). Thus the total predicted mass will grow if we take into account that low mass stars are removed due to mass segregation. This makes our statement on the IMF constrains even more significant.
In general, it is very hard to remove mass in form of stars and remnants from a nuclear cluster (a SMBH binary, however is able to remove stars efficiently). The processed gas mass, unlike stars and remnants, can be expelled from the nuclear cluster due to supernovae or AGN activity in the distant past of the SMBH. Stellar winds are also able to remove gas from the cluster. While the low mass main-sequence stars and giants have wind speeds significantly lower ($v_{\rm wind}\sim \rm few~ 10\,km/s$) than the escape velocity of the NSC the most massive stars ($v_{\rm wind}\sim \rm 1000\,km/s$) can efficiently remove gas from the cluster \citep{mclaughlin06,martins07}. Therefore, the processed gas mass cannot constrain the IMF although flat IMFs require a factor hundred greater gas masses than Kroupa like IMFs.  
Summarizing the arguments we can say that IMFs flatter than $\alpha > -1.1 $ violate the observed cluster mass. IMFs flatter than $\alpha > -0.8$ exceed even the total mass (SMBH+cluster).  

\subsubsection{Constraints from the diffuse H-band background}\label{sec:background}
\begin{figure}
\begin{center}
\includegraphics[width=0.8\columnwidth]{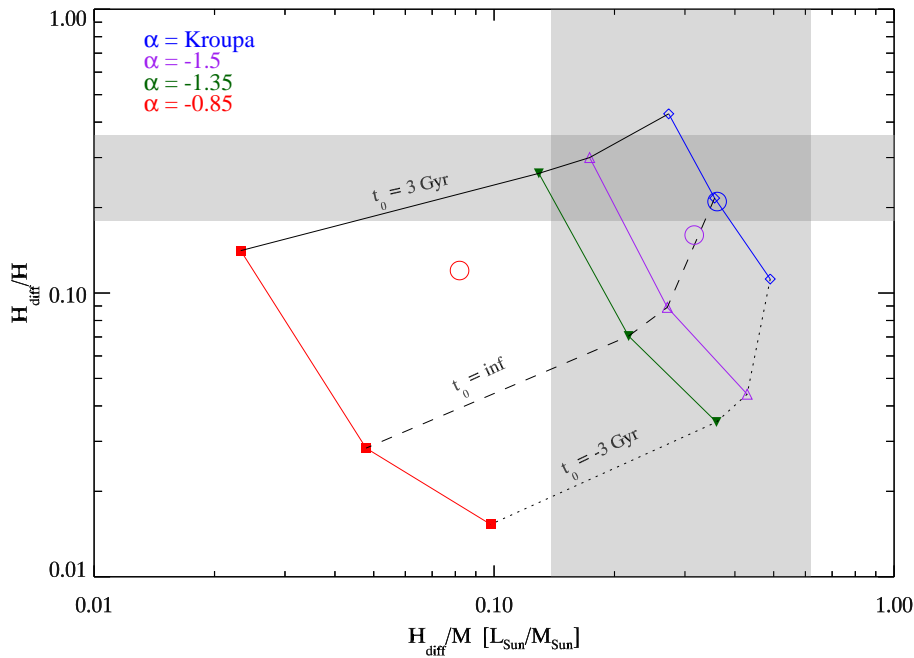}

\caption{The observed ratios of diffuse background light to the total H-band luminosity $H_{\rm diff}/H$ (horizontal bar shows 1\,$\sigma$ confidence interval) and the inverse mass-to-diffuse-light ratio $H_{\rm diff}/M_{\rm dyn}$ (vertical bar shows 1\,$\sigma$ confidence interval) constrain the IMF and the star formation history. The colors indicate the assumed IMF. The best-fit models (Tab.\ref{tab:SFR}) are represented by open circles. For comparison, simple models with exponentially declining, constant and exponentially rising (connected by solid, dashed, dotted lines) star formation are shown. The star formation rate in this case is parametrized as ${\rm SFR}(t)\propto e^{-t/t_0}$, where $t_0=\rm \inf$ corresponds to continuous star formation, $t_0=3\rm \,Gyr$ corresponds to a burst with an exponential decline timescale of 3\,Gyrs and $t_0=-3\rm \, Gyr$ corresponds to a raising star formation rate. The assumed cluster age is $t=13 \,\rm Gyr$. Models with an increasing star formation rate fail to produce the observed amount of background light, while models with decaying star formation underpredict the $H_{\rm diff}/M_{\rm dyn}$ ratio. The mismatch gets worse for flatter IMFs. Only the best-fit or the continuous model with a normal Kroupa IMF match both constraints.}
\label{fig:diffuse_background}
\end{center}
\end{figure}

As discussed in Sec.\,\ref{sec:diffuse_background}, depending on the IMF and star formation history the diffuse background light from faint unresolved stars may probe the bulk of the stellar population. We find that the diffuse background contributes $H_{\rm diff}/H=27\pm9\%$ to the total H-band flux. Furthermore, we derived a mass-to-diffuse-light ratio of $M_{\rm dyn}/H_{\rm diff}=2.6\pm1.5\, \msun/L_{H,\odot}$. In the following we used the inverse ratio $H_{\rm diff}/M_{\rm dyn}$ for convenience. We compared both quantities to our best-fit models (Sec.\,\ref{sec:fit}). To show how $H_{\rm diff}/H$ and $H_{\rm diff}/M$ depend on the IMF slope and overall star-formation history, we used the population synthesis code {\it STARS} \citep{sternberg98} to compute these ratios for a range of IMFs and star-formation time-scales $t_0$ for simple histories $SFR(t) \propto e^{-t/t_0}$. We considered a Kroupa IMF, and also three power-law IMFs, $dN/dm \propto m^{-\alpha}$, with $\alpha$ equal to -1.5, -1.35, and -0.85. All of the IMFs range from 0.01 to 120 M$_\odot$.  We used Geneva evolutionary tracks for solar metallicity stars, combined with empirical colors and bolometric corrections for dwarfs, giants, and supergiant stars. For low-mass stars ($< 0.8$~M$_\odot$ we computed the H-band luminosities along the lower main-sequence using the calibrations of Henry \& McCarthy (1993). We considered exponentially decaying ($t_0=3$ Gyr), continuous ($t_0=\infty$) and exponentially increasing ($t_0=-3$ Gyr) star-formation rates, for an assumed cluster age $t=13$ Gyr, and fixed dynamical mass $M_{\rm dyn}=1.5\times 10^6 \msun$.  The results are displayed in Figure\,\ref{fig:diffuse_background} in the $H_{\rm diff}/M$ versus $H_{\rm diff}/H$ plane. 
The gray areas indicate the measurement including the $1\,\sigma$ uncertainty. Figure \,\ref{fig:diffuse_background} shows that $H_{\rm diff}/M$ and $H_{\rm diff}/H$ both decrease for flatter IMFs. This behavior is due to the relative increase in remnant mass from massive stars, and the reduction in the relative fraction of diffuse light from low mass stars. For a given IMF, $H_{\rm diff}/M$ decreases but $H_{\rm diff}/H$ increases, as the history is altered from increasing, to steady, to declining star-formation, because the total luminosity for fixed mass is reduced for this sequence of histories, while the fraction of light produced by the accumulating long-lived low-mass stars is increased.  In Figure\,\ref{fig:diffuse_background} we also indicate the positions of the "best-fit" models discussed in Sec.\,\ref{sec:fit} and listed in Table\,\ref{tab:SFR}. Figure\,\ref{fig:diffuse_background} shows that the observed $H_{\rm diff}/M$ and $H_{\rm diff}/H$ ratios require steep (canonical) IMFs and continuous or mildly decaying star-formation. In particular, flat IMFs with $\alpha > -1.5$ are inconsistent with the observations\\

Our measurements of the diffuse H-band background can be compared to the findings of \cite{loeck10}, who used the diffuse K-band background together with the dynamical mass to derive $M/K_{\rm diff}=1.4^{+1.4}_{-0.7}\, \msun/L_{\odot,K}$, which was then compared with model predictions. Their mass-to-diffuse-light ratio is somewhat lower than our value ($M/K_{\rm diff}=1.9\pm1.2\, \msun/L_{\odot,K}$), yet consistent within the errors. The difference originates from the assumed extinction value. \cite{loeck10} used $A_K=3.3$ published by \cite{buchholz09}. However, this value is $0.5$ magnitudes (factor 1.6) higher than the most recent one published by \cite{schoedel10a,fritz11}. Correcting for this difference removes the discrepancy between their finding and ours. Their analysis favored an IMF steeper than $\alpha = -1.3$ together with a continuous star formation rate or an increasing star formation rate. Our analysis confirms their finding of a steep IMF. Taking into account the revised extinction our finding of an early star formation is also consistent with the findings of \cite{loeck10}. However, we additionally used the ratio of diffuse-light to total light to constrain the star formation history even further.

\section{Discussion}\label{sec:discussion}
All statements concerning the star formation history of the NSC have been made under the assumption that the traced population is representative for the whole cluster. The radial extent of the cluster is $\sim 5\, \rm pc$, i.e. larger than the radii probed with our SINFONI observations. However, the inner 1.2\,pc contain already $6\times10^6\,\msun$ of the total $30\times10^6\,\msun$. Thus we cover about 20\% of the total cluster mass and gives us confidence that our results are somewhat representative for the whole cluster. 
\subsection{Initial Mass Function in the GC}
We find that the old giant population must have formed with a  IMF steeper than $\alpha<-1.3$, probably with a normal Chabrier$/$Kroupa IMF. In any case, the IMF must have been significantly different from the IMF observed in the young stellar disks. \cite{paum06,bartko10} found that the young stellar disks in the GC have formed preferentially massive stars. They favored an almost flat IMF with $dN/dm\sim m^{-0.45}$ \citep{bartko10}. This indicates that the environmental conditions at the time of formation were different in the past. The nuclear cluster probably formed at a time, when the SMBH itself was younger and less massive. With the radius of influence smaller than today, the gravitational potential resembled more a normal cluster. Consequently the IMF was closer to the universal one. The nuclear cluster might also have been contaminated by in-spiraling clusters that had formed outside of the Galactic Center. Depending on their mass, clusters within several 10-100\,pc can spiral into the NSC during a Hubble time \citep{agarwal11}. Both effects lead to an IMF significantly steeper than the one observed in the young stellar disks today. 
\subsection{Star formation in the vicinity of the GC}
The Galactic Center region shows star formation at all ages. The massive Arches and Quintuplet clusters at distances of a few ten pc were formed 2 and 4\,Myrs ago \citep{figer02,figer99}. The stellar disk in the central cluster was formed about 6\,Myrs ago \citep{paum06,bartko09} and at roughly 50\,pc distance the Sgr\,B2 cloud harbors several massive star cluster in the making. \cite{figer04} used HST photometry of several fields within 100\,pc of the center to derive the star formation history. They find a continuous star formation in their FoV. \cite{genzel03} and \cite{buchholz09} used AO-assisted photometry of the nuclear cluster and found that the KLF is well matched by a bulge-like (8-10\,Gyr) population with an admixture of young main-sequence stars. \cite{blum03} used spectroscopic data of the most luminous giants within 2.5\,pc from Sgr\,A*. Our work confirms their favored formation scenario. The star formation happened predominantly at old times. An intermediate period showed a reduced star formation, while during the last few 100\,Myrs the star formation rate was increasing. This is also consistent with findings based on the stellar disks and the number of red supergiants. Roughly 75-90\% of the mass contained in the central cluster formed 5-12\,Gyrs ago. The formation of the cluster might coincide with the formation of the bulge around $10\pm2.5$\,Gyrs ago \citep{zoccali08}. The $\alpha$-element ratios require the bulge to have formed on a short timescale ($\rm \sim1\,Gyr$). Whether this is also true for the Galactic Center cannot be answered with the given uncertainty of the data. 
The fact that \cite{buchholz09} do not find a young giant population is not surprising, since those stars make up only $\sim10\%$ of the number counts. The KLF fitting is not sensitive enough and since the young giants are rather warm and bright, they might be attributed to the admixture of young main-sequence stars. The same is true if the results are compared to \cite{figer04}. They, however, find a continuous star formation rate within 100\,pc. This does not contradict our findings is consistent with the idea of recurrent episodic star formation in the greater Galactic Center region \citep{serabyn96}. The star formation events themselves happen on spatial scales of $\rm<1\,pc$ (e.g. Central-, Arches \& Quintuplet Cluster). If one assumes that the individual star formation events happen incoherently, then one would naturally expect a continuous star formation history on scales $\rm>>1\,pc$. Thus \cite{figer04} find a star formation history that is smoothed across many episodic single events.  

\section{Conclusion}\label{sec:conclusion}
We used several methods to constrain the star formation history of the Milky Way Nuclear Cluster. This is the only nuclear cluster in which individual stars can be resolved and reliable age estimates can be made. For this purpose, we used 450 K-band spectra of late-type giants to derive individual stellar temperatures. By using a CO index that is insensitive to systematic effects such as reddening or instrument resolution, together with the new stellar library of \cite{rayner09} we improved the temperature calibration for the red giants in the Galactic Center. Together with K-band photometry we were able to construct a detailed H-R diagram of the giant population. The comparison of the observed H-R diagram with model populations, allowed us to infer the star formation history of the Galactic Center.
Our results are as follows:
\begin{enumerate}
\item The bulk of the Nuclear Cluster is old. Roughly 80\% of the stellar mass formed more than 5\,Gyrs ago. It might have formed at the same time as the Galactic Bulge at a redshift of 1-2. 
\item After the bulk of the cluster had formed, a period of reduced star formation followed between 1-5\,Gyrs ago. The star formation reached a minimum $\sim 1\, \rm Gyr$ ago. Our inferred star formation history confirms earlier findings of \cite{blum03}. 
\item Only during the last 200-300\,Myrs star formation has set in at a significant level. A population of intermediate age giants are tracers of that period. Making up only 10\% of the number counts, the intermediate age population is hardly traceable with the KLF. However, they can be clearly identified in a H-R diagram due to their high temperatures. The spatial distribution and kinematics of those giants resemble the one of the old giant population.
\item We report the first detection of main-sequence B9/A0 stars with magnitudes $17<m_K<18$ in the GC. They are the faintest early-type stars having been found in the Galactic Center so far. With these stars we probe the mass regime between $2.2-2.8\,\msun$ and main-sequence lifetimes of 360-730\,Myrs. 
\item The ratio of late-type to early-type stars in the magnitude bin $17<m_K<18$ yields an independent estimate of the star formation rate during the last 500\,Myrs. We find that the average rate during the last 500\,Myrs must have been a factor 10 lower than the average rate between 1-12\,Gyrs ago. This finding supports the claim of an old cluster population.
\item We find that the bulk of the stellar mass must have formed with an IMF steeper than $dN/dm\sim m^{\alpha};~\alpha <-1.5$. Otherwise, the required remnant and stellar mass violates the observed dynamical mass and diffuse background. Thus, the bulk of the old stars formed with an IMF significantly steeper than the one observed in the young stellar disk. We suggest that this apparent discrepancy can be naturally explained if the stars formed at a time when the SMBH itself was much younger and less massive. Consequently the sphere of influence was significantly smaller. Without the extreme environment of a SMBH, the stars formed with an IMF close to the universal one. The SMBH was fed by gas including stellar mass loss \citep{freitag06} and dominated the inner few pc over time. With the growing sphere of influence, the IMF became flatter and reached the value observed in the young disk
\item The deep census of the Galactic Center using integral field spectroscopy yields a late-type KLF between $12<m_K<18$ with a slope of ${\rm d\, log}N/{\rm d}m_K=0.33\pm0.03$. This confirms the previous findings of \cite{genzel03,buchholz09}. The slope is consistent with an old Bulge like KLF \citep{alexander99}.  
\end{enumerate}
Taking into account systematic effects leads to an improved age estimate of the Nuclear Cluster. To improve the age estimate further, however, requires a technical leap. The main difficulty is that $\rm >5\,Gyrs$ old isochrones are spaced by only a few 10\,K. The necessary temperature accuracy can hardly be achieved by spectral analysis. The most promising way is to detect a turnoff of the main-sequence population. This requires the spectroscopic identification of stars with magnitudes of $m_K\approx19$. For example, the confirmation of late-type giants only 1 mag fainter than the current limit can unambiguously confirm the presence of stars with ages of $\rm>10\,Gyrs$. Yet, the required telescope time, which is needed for the spectroscopic identification of those stars, is tremendous. Furthermore, these faint stars are significantly affected by stellar crowding. Thus further progress will require the resolving power of a new generation of telescopes like the E-ELT or TMT. The same is true to derive a model-free measurement of the initial mass function in the vicinity of the SMBH.

\acknowledgements
This work has made use of the IAC-STAR Synthetic CMD computation code. IAC-STAR is supported and maintained by the computer division of the Instituto de Astrofisica de Canarias. AS thanks the DFG for support via German-Israeli Project
Cooperation grant STE1869/1-1.GE625/15-1. We thank the referee for his valuable comments.

 \bibliographystyle{apj}
\bibliography{ms}

\end{document}